\documentclass[reprint,
superscriptaddress,
 amsmath,amssymb,
 aps,
 prd,
floatfix,
]{revtex4-2}

\usepackage{graphicx}
\usepackage{dcolumn}
\usepackage{bm}
\usepackage{hyperref}
\usepackage{aas_macros}
\usepackage{textcomp}
\usepackage{xfrac}
\usepackage{comment}
\usepackage{multirow}   

\usepackage{makecell}

\usepackage[font=small,labelfont=bf,
   justification=justified,
   format=plain]{caption}
\usepackage{xcolor}

\begin{document}

\preprint{arXiv:xxxx.xxxx}

\title{Superhorizon isocurvature fluctuations relax tensions}

\author{Alessandra Fumagalli}
\affiliation{Dipartimento di Fisica - Sezione di Astronomia, Universit\'a di Trieste, Via Tiepolo 11, 34131 Trieste, Italy}
\affiliation{INAF-Osservatorio Astronomico di Trieste, Via Tiepolo 11, I-34131 Trieste, Italy}
\affiliation{IFPU, Institute for Fundamental Physics of the Universe, via Beirut 2, 34151 Trieste, Italy}
\affiliation{INFN, Sezione di Trieste, Via Valerio 2, 34127 Trieste, Italy}

 \author{Yodovina Pi\v{s}kur}
 \affiliation{SISGA, Via dei Catterini 27, 34170 Gorizia, Italy}

\author{An\v{z}e Slosar}
 \affiliation{Physics Department, Brookhaven National Laboratory, Upton, New York 11973}

\date{\today}

\begin{abstract}
We present a new class of models that have potential to alleviate tensions present in the cosmological data today.
We postulate about variation in the sound horizon scale on superhorizon scales, i.e., on scales that are larger than that of the present observable low-redshift Universe ($\gtrsim 1\,$Gpc) while at the same time smaller than the largest scales probed by the cosmic microwave background (CMB) ($\lesssim10\,$Gpc). In this scenario, CMB peaks are naturally smoothed as preferred by the Planck data, while at the same time the low-redshift baryon acoustic oscillation calibration is partially decoupled from the CMB. Taking superhorizon variations in baryon fraction as an example and using approximate modeling, we find improvement in the best fit Planck power spectrum model $\Delta \chi^2 \sim 6$ for 1 extra degree of freedom with the relevant extension parameter $10^3 \sigma_b = 2.10 \pm 0.60 $, implying about 10\% variations in baryon fraction across the Universe.   At the same time, $S_8$ drops by about 1 sigma, easing tension with weak lensing surveys. We find that the combination of Planck 2018 data, eBOSS BAO data, and Riess \textit{et al.} distance ladder Hubble parameter determination produce a satisfactory fit in our model if we allow for a phantom dark energy equation of state.
\end{abstract}

\maketitle

\newcommand{\ob}{\omega_{b}}
\newcommand{\obl}{\omega_{b,L}}
\newcommand{\sob}{\sigma_{b}}

\newcommand{\obc}{\omega_{b,C}}
\newcommand{\sobc}{\sigma_{b,C}}

\section{Introduction}

The Hubble tension is currently perhaps the most intriguing tension in cosmology. It is a tension in measurements of the expansion rate of the Universe when performed directly, using a cosmological distance ladder, and that inferred indirectly from the measurements of the high redshift Universe, most notably the cosmic microwave background (CMB). The recent measurements using distance ladder are $H_0 = 73.04\pm1.04$ \cite{2112.04510}, while Planck 2018 data fit for a flat $\Lambda$CDM model find $H_0=67.31\pm0.59$ \cite{1807.06209}.  There have been many attempts at reconciling the two measurements in extended cosmologies without any convincing solution \cite{1908.03663, 2107.10291,2103.01183}. 

The difficulty is illustrated by the inverse distance ladder argument. In this measurement, we take baryonic acoustic oscillations (BAO) measurement of the Hubble parameter at redshift $z\sim 0.5$ and transport it to $z=0$ using supernovae Ia distance measurement \cite{1411.1074,1411.1094}. This can be done in an essentially model-free manner, assuming any sufficiently flexible smooth evolution of background cosmology. If we then calibrate the BAO distance scale using either CMB or standard pre-recombination physics with a prior on baryon density, we recover the Planck preferred value of $H_0$. This implies that the solution \emph{must} reside in a prerecombination physics. However, any change in prerecombination physics would generically affect the damping tail of CMB, rendering such solutions either unlikely or very fine-tuned \cite{1908.03663}.

Here we propose a different solution to the problem, by postulating that CMB is subject to a locally different cosmology. We do this by imposing large scale fluctuations in the sound horizon with sufficiently large correlation lengths that our observable, low-redshift Universe appears essentially homogeneous, while the CMB surface contains several patches. This is not difficult to arrange since the comoving distance to the redshift $z=6$ is about 8\,Gpc, while the comoving distance to the surface of last scattering at $z\sim 1150$ is about 13.6\,Gpc. A large-scale field with a correlation length of $\sim 10$\,Gpc will be approximately constant across the observable Universe, while presenting $O(10)$ uncorrelated patches across the CMB surface. This way we decouple the sound horizon inferred from the CMB peaks from that present in the local BAO measurements and thus weaken the inverse distance ladder argument. 

In order to put such model on a concrete footing, we employ very large scale baryon fraction fluctuations as means to achieve sound horizon fluctuations. We do this by varying physical baryon density $\omega_b = \Omega_b h^2$ at fixed matter density fluctuations, so that the background cosmology remains simple and unchanged. Technically, these are called the compensated isocurvature fluctuations, since the baryon density fluctuations are compensated by dark matter fluctuations, while leaving radiation fields untouched. They have been studied before \cite{0907.3919,1306.4319,1704.03461} in other contexts. They leave no imprint on CMB at linear order, since both the equation of state of the baryon-dark matter fluid and then potential fluctuation are unchanged.  The main novelty in our paper is that we consider a cutoff in the power spectrum of these fluctuations on very large scales which leaves the small scale physics unchanged, and to consider large, beyond linear-order amplitudes.

As a side effect, one expects the CMB power spectrum to be somewhat different in patches of CMB with a different value of the field, leading to a smearing of the acoustic peaks in the CMB, which is not present at linear order. Such smearing of CMB peaks has in fact been observed and manifests itself as a preference for CMB data alone for a high value of lensing amplitude parameter $A_L$ or equivalently closed Universe. This is a known fact that has been discussed and explained at length in Planck 2015 and 2018 papers \cite{1502.01589, 1807.06209} and again pointed in more recent papers \cite{1911.02087,1908.09139}.

In this paper, we fit Planck 2018 data using an approximate methodology which should nevertheless capture all the essential physics at play.  
As expected, we find improvement in the $\chi^2$ that, while not as pronounced as in the case of the $A_L$ and $\Omega_k$ parameters, is significant and formally a 4 sigma preference for a nonzero value of our parameter.  At the same time, we find shifts and expanded errorbars in some other parameters. In particular, we find that the value of the $S_8$ parameter drops and that its error increases. This completely alleviates tension with the amplitude of fluctuations, which seems to be low in the low-redshift experiments \cite{1809.09148,2105.13549,2105.13543,2105.13544,2204.02396,1502.01597,2105.03421,2206.10824,2007.15632,2105.12108}.
We conclude with discussion of strengths and weakness of this model in the final section.

\section{Modeling}
As described in the previous section, we promote the physical baryon fraction parameter $\omega_b = \Omega_b h^2$ to be a position dependent field $\omega_b(\mathbf{x})$. We use standard notation where $\Omega_b$ is the baryon density expressed as a fraction of critical density in the Universe and  $h=H_0/(100\,{\rm km}\,s{\rm }^{-1}{\rm Mpc}^{-1})$ is the Hubble parameter. So defined baryon density parameter is proportional to the physical baryon density today, i.e. one measured in kg/m$^3$. We assume field $\omega_b$ is a slowly varying field across the Universe with the correlation length of around $r_c\sim 10\,$Gpc. 
Therefore, any observer will measure the local baryon density $\obl$ to be drawn from  a Gaussian distribution
\begin{equation}
    P(\obl | \ob, \sob) = \frac{1}{\sqrt{2\pi \sob^2}}\exp\left[ -\frac{(\obl - \ob)^2}{2\sob^2}\right],
\label{eq:obl}
\end{equation}
and its value will be approximately constant for separations  $r \ll r_c$ and uncorrelated for separations $r \gg r_c$. The new parameter $\sigma_b$ crucially determined how much baryon fraction is allowed to vary across the Universe.

Importantly, we do this at a constant overall matter density, $\omega_{\rm m}$, so any increase in baryon density is compensated for by an appropriate decrease in the dark matter density and the overall density and their gravitational fluctuations follow the standard cosmology.  In other words, we have introduced large scale isocurvature fluctuations. Most importantly, the background cosmology is unchanged.

Since we have imposed these fluctuations on superhorizon scales, each $r_c$ sized patch of the Universe evolves as in a standard cosmology. Since 10\,Gpc corresponds to redshift $z\sim 20$, the entirety of our local Universe will appear as a standard $\Lambda$CDM Universe with a certain $\obl$. On the other hand, since comoving distance to CMB is about 13\,Gpc,  the CMB surface will contain approximately $N_p=10-20$ independent patches, each with an independent value of baryon fraction.

\subsection{CMB power spectrum}

\newcommand{\mC}{\rm C}
\newcommand{\mCell}{{\rm C}_{\ell \ell'}}
\newcommand{\hn}{\hat{n}}
To build an intuition of how the CMB power spectrum changes, we take a somewhat nonstandard approach by starting with a white-noise random field on the sphere that we filter into CMB fluctuations in a position dependent manner. We therefore write CMB temperature fluctuation in a direction $\hn$ as
\begin{equation}
    \Delta T (\hn) = \sum_{\ell,m} a_{\ell m } R_\ell(\hn)  Y_{\ell m} (\hn),
\end{equation}
where $Y_{\ell m}$ are spherical harmonics, $a_{\ell m}$ is a unit variance zero mean Gaussian random field satisfying $a_{\ell m} a^*_{\ell' m'} = \delta^K_{\ell \ell'} \delta^K_{m m'}$ and the transfer function $R_\ell$ satisfies $ R_\ell^2(\hn) = C_\ell(\hn)$.  In a standard cosmology $R_\ell$ would be position independent, but now it corresponds to a power spectrum in the direction $\hn$, which is modulated because the baryon fraction is modulated. After some trivial manipulation, we find that the correlation between two points on the sky is given by
\begin{equation}
    \left< \Delta T(\hn) \Delta T(\hn') \right> = \sum_{\ell m} (2\ell+1) \sqrt{C_\ell (\hn) C_\ell (\hn')} P_\ell(\theta)
\end{equation}
where $\hn \cdot \hn' = \cos \theta$. Owing to the position dependent power spectrum, the correlation between any two points does not depend only on their separation, but is instead given by the correlation function in the Universe whose power spectrum is the geometric mean of the power spectra in the two directions of the sky of interest.  

When we calculate the full sky correlation function in such a Universe, we get a pair weighted geometric average of the power spectra in different directions of the sky. On small separations, smaller than the scale corresponding to $r_c$,  both legs of the correlation function fall onto the same patch.  Since our $r_c$ corresponds to $\ell \sim 10$, essentially the entire power spectrum falls in that regime. And since $\omega_b$ does not change the large-scale CMB power spectrum, the resulting power spectrum and its covariance in a single patch is simply
\begin{eqnarray}
\left< C_\ell^{\rm patch} \right> &=& \frac{1}{2\pi \sigma_b} \int C_\ell (\obl) P(\obl|\ob,\sigma_b) d\obl \label{eq:mean}\\
 \mC^{\rm patch}_{\ell \ell'} &=& \left< C_\ell C_\ell'\right> - \left<C_\ell (\ob)\right> \left<C_{\ell'} (\ob)\right>  \nonumber \\
 &=& \frac{1}{2\pi \sigma_b} \int C_\ell (\obl) C_{\ell'} (\obl) d\obl P(\obl|\ob,\sigma_b) \nonumber \\
 &&-\left<C_\ell (\ob)\right> \left<C_{\ell'} (\ob)\right> \,,\label{eq:cov}
\end{eqnarray}
where we consider $N_p$ patches, each generated with $C_\ell (\obl)$ with $\obl$ drawn from a Gaussian $\ob$ and variance $\sigma_b$.

Here we see two distinct effects at play. First, the expected mean power spectrum changes, smearing peaks over values of local baryon density $\obl$. But importantly, the local power spectrum in any one patch will suffer a large sample variance, since we do not know its local value of $\obl$, which is in general not equal to the mean global baryon density. Over $N_p$ patches in the sky, this covariance decreases to ${\mC}^{\rm patch}_{\ell \ell'}/N_p$, but unless $N_p$ is very large it cannot be ignored.

For small enough $\sigma_b$, we can rely on Taylor expansion:
\begin{equation}
    C_\ell (\obl) = C_\ell (\ob) +  \frac{\partial C_\ell}{\partial \ob} \Delta \ob + \frac{1}{2}\frac {\partial^2 C_\ell}{\partial \ob^2} \Delta \ob^2 +\ldots
\end{equation}
Then to the leading order, Eqs. \eqref{eq:mean} and \eqref{eq:cov} simplify to
\begin{eqnarray}
\left< C_\ell \right> &=& C_\ell (\ob) + \frac {\partial^2 C_\ell}{\partial \ob^2} \frac{\sigma_b^2}{2} + O(\sigma_b^4) \label{eq:mean2}\,,\\
 \mC_{\ell \ell'} &=& \frac{\partial C_\ell}{\partial \ob} \frac{\partial C_\ell'}{\partial \ob} \frac{\sigma_b^2}{N_p}
 +O(\sigma_b^4)\,, \label{eq:cov2}
 \end{eqnarray}
 where now $ \mC_{\ell \ell'}$ is the full observed sky covariance.
 
In the absence of a Boltzmann code with numerical differentiation, we resort to performing numerical derivative using a 3-point stencil. Setting the stencil step size to be $\sigma_b$, Eq. \eqref{eq:mean2} simplifies again, giving
\begin{equation}
\left< C_\ell \right> \approx \frac{1}{2} \left( C_\ell (\ob+\sigma_b) + C_\ell(\ob-\sigma_b) \right) + O(\sigma_b^4)\,. \label{eq:mean3}
\end{equation}
This simple prescription is just the mean of two CMB power spectra evaluated at values of baryon fraction shifted by $\sigma_b$.
We have tested that this expression is a few percentage level accurate in the \emph{correction} to the power spectrum compared to the numerical integration of the full integral \ref{eq:mean} up to $\sigma_b=0.005$. Since the correction is in turn at the percentage level of the power spectrum, this is sufficient precision given our constraints.

In principle, one should add ${\mC}_{\ell\ell'}$ to the power spectrum covariance matrix. However, as written in Eq. \eqref{eq:cov2}, it is clear that this contribution has a single eigenvector corresponding to the first derivative of the power spectrum with respect to $\ob$. Therefore, we can simply ignore this term, as long as we are cognizant that our $\ob$ will measure the baryon density averaged over the CMB patches ($\obc$), rather than the global mean baryon density. This difference is taken into account when calculating the local $\obl$ in the next section.

\subsection{Baryonic Acoustic Oscillations}

To consistently apply likelihood for Baryonic Acoustic Oscillations, we need to be able to correctly set the value of sound horizon or more precisely $r_d = r_d (\obl, \omega_{\rm m}, N_{\rm eff})$. 
To get probability for the local value of baryon density $\obl$, we must marginalize over possible values for the global values:
\begin{equation}
    P(\obl) = \iint P(\obl | \ob, \sob) P(\ob, \sob) d\ob d\sob\,.
\end{equation}
What we measure are the mean and variance over the $N_p$ patches present in the CMB. Writing down the problem as a general problem of probability of drawing a number from a Gaussian distribution for which we have $N_p$ measurements with \emph{sample} mean $\obc$ and \emph{sample} variance $\sobc$, we can derive the expression of the probability of $P(\obl | \obc, \sobc)$. The full expression is given in the Appendix \ref{app:pdf}. 

\begin{figure}
    \includegraphics[width=\linewidth]{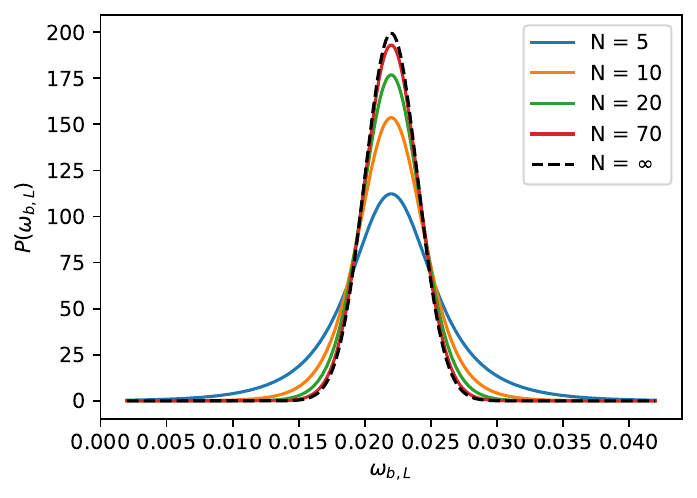}
    \caption{Example of $P(\obl)$ as a function of number of patches. In this case we fixed $\obc=0.022$ and $\sobc=0.002$, but in reality these numbers are themselves uncertain. In the limit of $N_p\rightarrow \infty$, we expect the mean and variance measured from the patches to be perfectly determined, in which case $P(\obl|\obc,\sobc) = G(\obl; \obc,\sobc)$  plotted as black dashed line here. \label{fig:smallvar}}
\end{figure}

The behavior of this function is shown in the Fig. \ref{fig:smallvar}.  For BAO measurement, we therefore additionally measure the local baryon density $\obl$ as a free parameter, which sets the calibration for the local BAO measurement and is subject to prior in Eq. \eqref{eq:oblprior}. In our inference problem, $\sobc$ and $\obc$ are themselves determined only up to an error. So, in every step in a Monte Carlo Markov Chains (MCMC) chain, the current values of $\obc$ and $\sobc$ are used to determine the prior on $\obl$ and $\obl$ is in turn used to determine the local sound horizon to be used in BAO likelihood. The width of $P(\obl)$ depends on the number of patches $N_p$ in the CMB. The lesser $N_p$, the less well the global mean and variance are determined and the wider the distribution. In this work we use the $N_p=10$, for which the width of the distribution is about 50\% larger compared to $N_p = \infty$ limit. However, our results do not sensitively depend on this parameter.

\begin{figure*}
    \centering
    \includegraphics[width=0.9\textwidth]{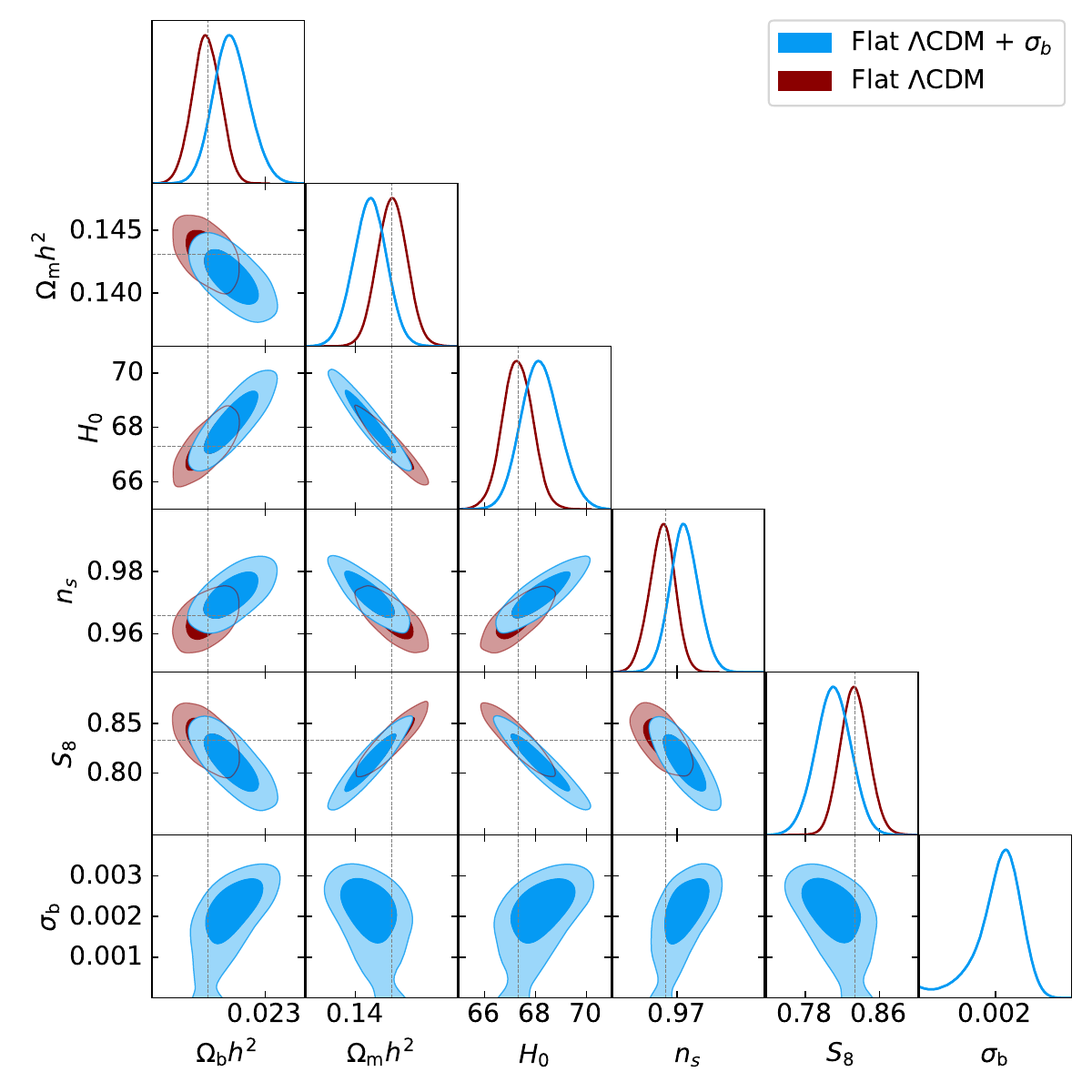}
    \caption{Corner plot for fits to Planck 2018 in the standard cosmology and one with large-scale baryon isocurvature fluctuations.}
    \label{fig:triangle}
\end{figure*}

\section{Results} \label{sec:results}
Our baseline fits use the Planck 2018 data, where we fit an additional $\sigma_b$ parameter. We use the Cobaya \cite{Torrado:2020dgo} fitting system for evaluating the likelihood function, and we infer the cosmological parameters with a MCMC approach, by using the python wrapper for the nested sampling PolyChord \cite{Handley:2015fda}. We consider the  baseline TT,TE,EE+lowE likelihood, modified to take as input the mean $C_\ell$ of Eq.~\ref{eq:mean3}; the CMB power spectra are calculated by CAMB \cite{Lewis:1999bs}. We adopt the same priors and nuisance parameters of the Planck 2018 analysis. This means that our flat-prior parameters are $\omega_b$, $\omega_{\rm m}$, $n_s$, $\log(10^{10} A_s)$, $\theta_{\rm MC}$, and $\tau$. For precise definitions of these parameters, see \cite{Lewis:1999bs}.

For the BAO, we analyze the BOSS DR12 data \cite{BOSS:2016wmc} by adopting the  ``consensus'' likelihood described in Planck 2018, also modified to take into account the local value of the baryon density parameter, namely $\omega_{b,L}$. The prior distribution for the latter is set by Eq.~\eqref{eq:oblprior}.  The sound horizon is calculated as a function of $\obl$ and $\obc$ using a fitting function from \cite{1411.1074} shown to be sufficiently accurate for eBOSS data. When we add $H_0$ prior from distance ladder, we simply importance sample these results with a Gaussian prior on $H_0$.

We are also interested in the best fit $\chi^2$ values. We derive these in two ways: by taking the lowest $\chi^2$ in the MCMC chain, or by explicitly running the Cobaya supplied minimizer. MCMC chains are likely less good at finding the precise minimum, but also much more robust against falling into the local minima. In general, we find consistent results, and we take the difference as a measure of uncertainty.

We start by fitting $\Lambda$CDM with a free $\sigma_b$ parameter. We show our results in the triangle plot in Fig. \ref{fig:triangle} and Table \ref{tab:main}. The best fit $\chi^2$ improves by $\sim 5.3$ units, leading to an approximately 2.3 sigma frequentist preference. The marginalized posterior for $10^3 \sigma_b = 2.10 \pm 0.060$, which corresponds to 4-sigma preference for the nonzero value of this parameter.  We see that several parameters have increased variance, which we discuss in conclusions.

\begin{figure}
    \centering
    \includegraphics[width=\linewidth]{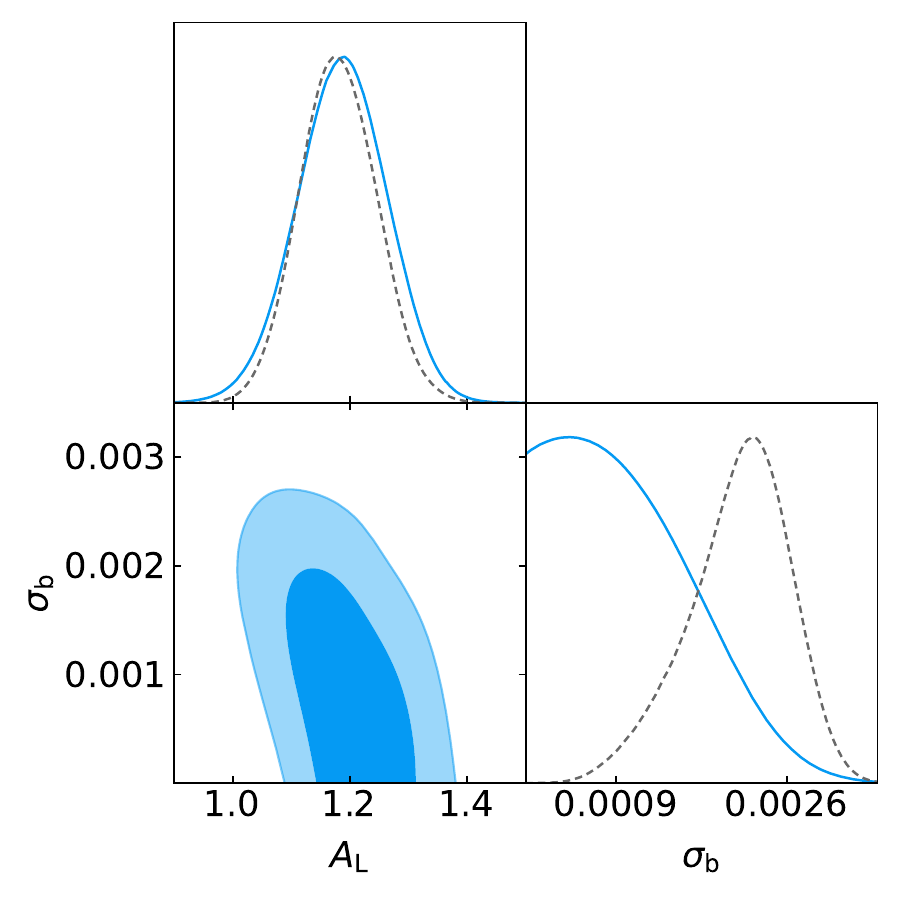}
    \includegraphics[width=\linewidth]{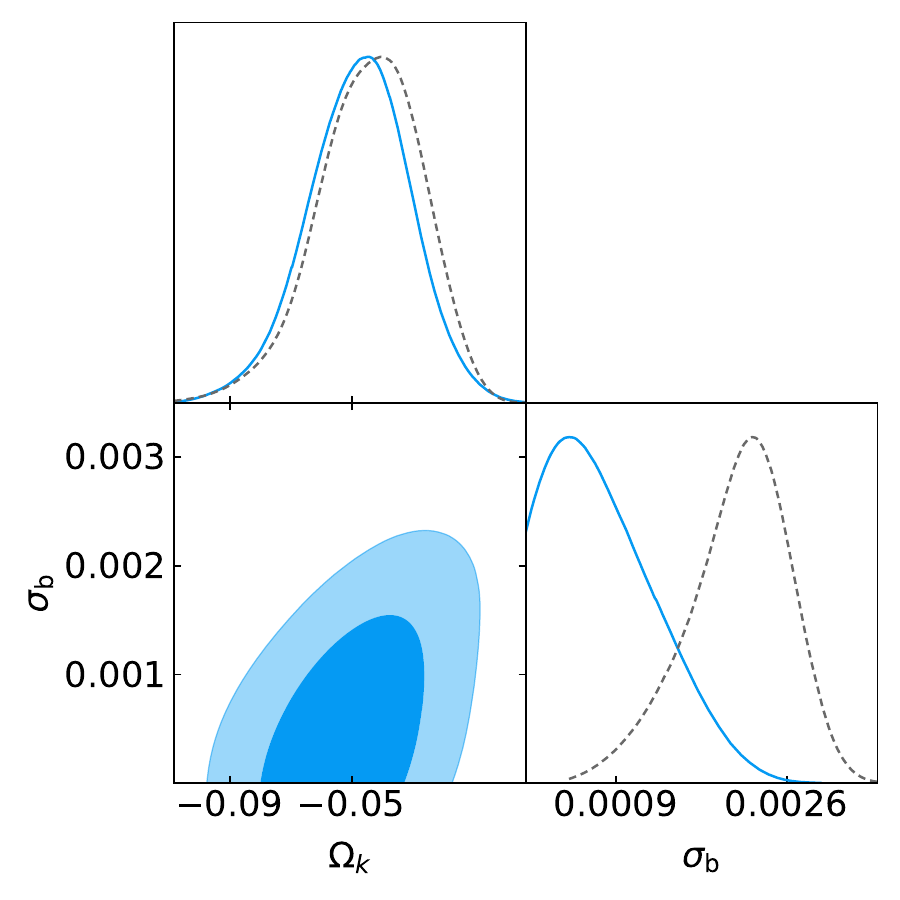}
    \caption{The posterior distribution when $\sigma_b$ is free in addition to either $A_L$ (top plot) or $\Omega_k$ (bottom plot). The parameters are degenerate, but not as strongly as might be expected. Gray dashed lines correspond to posteriors before addition of the other parameter in the plot.}
    \label{fig:add}
\end{figure}

We additionally run chains with freeing one additional parameter in addition to $\sigma_b$: $A_L$, $\Omega_k$, $N_{\rm eff}$ or sum of neutrino masses. The expectation was that in runs with $A_L$ and $\Omega_k$ the values would be heavily degenerate with $\sigma_b$, but this is only partially the case as illustrated in the Fig. \ref{fig:add}.
In this case, the best-fit model does not improve with respect to free $\sigma_b$ case, but we find only marginally changed constraints.

\begin{figure}
    \centering
    \includegraphics[width=\linewidth]{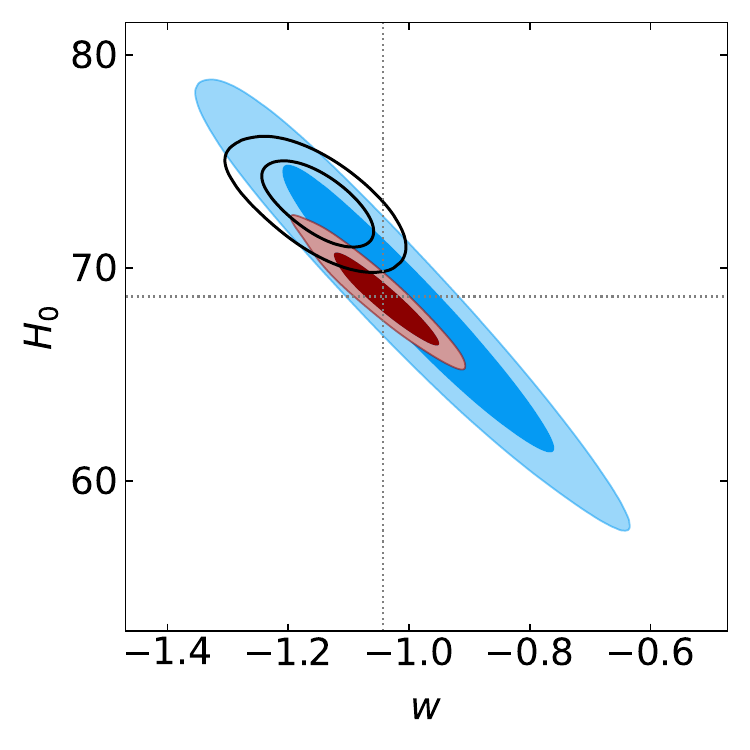}
    \caption{Posterior on the $H_0-w$ plane in the wCDM model. Red filled contours correspond to the baseline Planck2018 fit. Blue filled contours are the fit with a free $\sigma_b$ and free local baryon fraction. Black contours correspond to the addition of the distance ladder prior on $H_0$. }
    \label{fig:w}
\end{figure}

Finally, we also run a chain for the wCDM model. This is an interesting model: in the standard cosmology Planck 2018 is unable to constrain it as $w$ is almost perfectly degenerate with other parameters on the geometrical degeneracy. However, upon the addition of BAO, the degeneracy is broken and one immediately recovers the standard LCDM with $w$ strongly constrained around the cosmological constant value. This combination is in strong tension with the low-redshift measurements of the Hubble parameter. In our case, owing to decoupling of the local baryon density from the global one, the BAO is much less constraining.  This is illustrated in the Fig. \ref{fig:w}. Of course, the low-redshift sound horizon is not completely free, since the local baryon density must still be consistent with the distribution partially constrained by the CMB peak smearing, but the degeneracy is only weakly broken. We find that $H_0=68.3 \pm 4.0$ compared to  $H_0 = 68.7 \pm 1.5$ in the standard cosmological case. The $H_0$ errorbar has relaxed by a factor of 3, allowing one to raise the possibility of distance ladder consistency.  To check this possibility, we importance sample using the results of \cite{2112.04510}. In this case, we find that that best fit $\chi^2$ has not increased.   We thus have a model that fits CMB better than the baseline $\Lambda$CDM \emph{and} is consistent with both BAO and the distance ladder measurement. However, this comes at the price of a phantom dark energy with $w=-1.15\pm 0.054$. 

In Appendix \ref{app:triangles} we show additional full triangle plots. 

\begin{table*}[t]
\centering           
    \scalebox{0.92}{
    \begin{tabular}{l l  c c  c c }   
    \hline
    Model & Parameter & Base model & +\,$\sigma_b$ & Shift in value & Shift in error\\
      \hline
    $\Lambda$CDM & $\Omega_{\rm b}\,h^2$ & $0.02236 \pm 0.00015$  & $0.02263 \pm 0.00018$ & $+1.15\, \sigma$ & $\times\, 1.19$\\
    & $\Omega_{\rm m}\,h^2$              & $0.14316 \pm 0.00126$  & $0.14131 \pm 0.00137$ & $-1.01\, \sigma$ & $\times\, 1.07$\\
    & $H_0$                              & $67.29922 \pm 0.58864$  & $68.1815 \pm 0.72594$ & $+0.97\, \sigma$ & $\times\, 1.21$ \\
    & $n_s$                              & $0.96483 \pm 0.00441$  & $0.97219 \pm 0.00468$ & $+1.19\, \sigma$ & $\times\, 1.06$ \\
    & $S_8$                              & $0.83307 \pm 0.01551$  & $0.81018 \pm 0.01871$ & $-0.94\, \sigma$ & $\times\, 1.20$ \\
    & $10^3\sigma_b$                     & ...                     & $2.10    \pm 0.60$    &      ...          &        ...      \\
    
    \quad $\chi^2$ ($\chi^2_{\rm min}$) &                     & $2765.19$ ($2765.77$)    & $2759.92$ ($2759.84$)    & $-5.27$ ($-5.93$) &   \\
    \quad Evidence ratio                 &                         &                       & $4.26$          \\  
   \hline
   
   $\Lambda$CDM\,+\,$\Omega_k$  & $\Omega_k$ & $-0.0435 \pm 0.0171$ & $-0.0484 \pm 0.0182$ & $-0.19\, \sigma$ & $\times\, 1.06$  \\
    & $10^3\sigma_b$                     & ...                     & $0.99 \pm 0.62$    &      ...          &        ...      \\
    \quad $\chi^2$ ($\chi^2_{\rm min}$)&                     & $2756.67$ ($2755.57$)     & $2753.88$ ($2755.98$ )   & $-2.79$ ($+0.41$) &   \\
    \quad Evidence ratio                 &            &                      & $0.34$               &                                     \\  
    \hline
    
    $\Lambda$CDM\,+\,$A_L$ & $A_L$  & $1.180 \pm 0.065$ & $1.185 \pm 0.075$ & $-0.06\, \sigma$ & $\times\, 1.15$  \\
    & $10^3\sigma_b$                     & ...                     & $1.30 \pm 0.55$    &      ...          &        ...      \\
    \quad $\chi^2$ ($\chi^2_{\rm min}$ ) &                     & $2757.26$ ($2755.47$)    & $2755.85$ ($2755.86$)   & $-1.41$ ($+0.39$) &   \\
    \quad Evidence ratio                 &        &                   & $0.40$            &                                     \\  
    \hline
    
    $\Lambda$CDM\,+\,$N_{\rm eff}$ & $N_{\rm eff}$  & $2.918 \pm 0.187$ & $2.835 \pm 0.156$ & $-0.34\, \sigma$ & $\times\, 0.83$  \\
    & $10^3\sigma_b$                     & ...                     & $2.15 \pm 0.55$    &      ...          &        ...      \\
    \quad $\chi^2$ ($\chi^2_{\rm min}$)&                     & $2764.70$ ($2765.98$)    & $2758.67$ ($2760.07$)   & $-6.03$ ($-5.91$) &   \\
    \quad Evidence ratio                 &                &                   & $7.91$            &                                     \\  
    \hline
    
    $\Lambda$CDM\,+\,$m_\nu$ & $m_\nu$  & $0.091 \pm 0.0834$ & $0.139 \pm 0.136$ & $0.30\, \sigma$ & $\times\, 1.63$  \\
    & $10^3\sigma_b$                     & ...                     & $2.24 \pm 0.52$    &      ...          &        ...      \\
    \quad $\chi^2$ ($\chi^2_{\rm min}$)&                     & $2762.54$ ($2766.24$)    & $2759.76$ ($2759.88$)   & $-2.78$ ($-6.36$)&   \\
    \quad Evidence ratio                 &          &                    & $24.0$            &                                    \\  
    \hline
    
    $w$CDM (w BOSS BAO) & $w$  & $-1.04320 \pm 0.05828$ & $-0.99290 \pm 0.14521$ & $0.32\, \sigma$  & $\times\, 2.49$ \\
    & $\omega_{b}$             & $0.02238  \pm 0.00015$ & $0.02269  \pm 0.00028$ & $1.00\, \sigma$  & $\times\, 1.88$ \\
    & $\omega_{b,{\rm loc}}$   &  ...                    & $0.02202  \pm 0.00350$ & ...               & ...              \\
    & $r_d$                    & $147.1013 \pm 0.2801$  & $147.207  \pm 0.476$   & $0.19\, \sigma$  & $\times\, 1.70$ \\
    & $r_{d,{\rm loc}}$        & ...                     & $147.743  \pm 7.528$   & ...               & ...              \\
    & $H_0$                    & $68.6693  \pm 1.4601$  & $68.22466 \pm 4.31130$ & $-0.10\, \sigma$ & $\times\, 2.95$ \\
    & $S_8$                    & $0.82759  \pm 0.01272$ & $0.80535  \pm 0.02760$ & $-0.73\, \sigma$ & $\times\, 2.17$ \\
    & $10^3\sigma_b$                     & ...                     & $2.11 \pm  0.79$    &      ...          &        ...      \\

    \quad$\chi^2$ ($\chi^2_{\rm min}$)           &           & $2770.87$ ($2769.41$)    & $2772.95$ ($2762.48$)    & $+2.08$ ($-6.93$) &   \\
    \quad Evidence ratio       &                        &                        & $2.45$           &                 \\  
    \hline\\[-3.0ex]
    \makecell{$w$CDM (w BOSS BAO) \\+ local $H_0$} &  $w$  & $-1.16203 \pm 0.03983$ & $-1.15091 \pm 0.05471$ & $0.16\, \sigma$  & $\times\, 1.37$ \\
    & $\omega_{b}$             & $0.02236 \pm 0.00015$  & $0.02268 \pm 0.00027$ & $1.04\, \sigma$  & $\times\, 1.80$ \\
    & $\omega_{b,{\rm loc}}$   &  ...                    & $0.02453 \pm 0.00229$ & ...               & ...              \\
    & $r_d$                   & $146.8916 \pm 0.25557$  & $147.132 \pm 0.453$   & $0.46\, \sigma$  & $\times\, 1.77$ \\
    & $r_{d,{\rm loc}}$       & ...                      & $145.637 \pm 4.303$   & ...               & ...              \\
    & $H_0$                   & $71.8463 \pm 0.9468$    & $73.0434 \pm 1.1699$  & $0.79\, \sigma$  & $\times\, 1.24$ \\
    & $S_8$                   & $0.82757 \pm 0.01201$   & $0.79594 \pm 0.02540$ & $-1.12\, \sigma$ & $\times\, 2.11$ \\
    & $10^3\sigma_b$                     & ...                     & $2.13 \pm 0.83$    &      ...          &        ...      \\
    \quad$\chi^2$ ($\chi^2_{\rm min}$)           &           & $2779.17$  ($2772.15$)    & $2773.02$ ($2766.58$)   & $-6.15$ ($-5.57$ )&   \\
    \quad  Evidence ratio         &                            &                        & $1.83$                 &                                    \\  
    \hline
    \end{tabular}}
    \caption{Comparison of parameter fits in standard cosmology and one with large-scale baryon isocurvature fluctuation, using the baseline TT,TE,EE+lowE likelihood and, in the $w$CDM case only, BAO information. In addition to the shown parameters, all standard cosmological parameters are free to vary as discussed in Sect. \ref{sec:results}.  The $\chi^2$ row shows the minimum $\chi^2$ derived from the MCMC chains and from minimization. Evidence ratios were computed from chains using a nested model approach with extended model prior $\sigma_b\in[0,0.005]$). When adding BAO and $H_0$ information, the effective number of degrees of freedom changes. 
    \label{tab:main}  }
\end{table*}

\section{Discussion and Conclusions}
\label{sec:conclusions}

We have presented a new class of models that can relieve the internal tension in Planck and loose the BAO's iron grip on the $H_0$ tension. These models rely on superhorizon fluctuations in any parameter that can modulate the size of the sound horizon. We trust that creative theorists are able to build inflationary models with such behavior.

We have illustrated this by considering baryon fraction fluctuations across the Universe. When fitting the Planck 2018 data, we find $\sigma_b = (2.1 \pm 0.5)\times 10^{-3}$. However, owing to high non-Gaussian posterior shape, the model $\sigma_b>0$ is model preferred at the level of $\Delta \chi^2 \sim 6$ or Evidence ratio around $\sim 4$. Our limits are consistent with much tighter limits on the compensated isocurvature modes considered in \cite{1306.4319} and \cite{1704.03461}, because we consider modes that exist only on very large, superhorizon scales.


When fit with either free $A_L$ or $\Omega_k$, our model does not seem to provide a better fit, since those parameters are still better at fitting the residuals. 


\emph{Implications for $H_0$ tension.}
We find that simply adding the $\sigma_b$ parameter raises the inferred value of $H_0$ by about 1 sigma and increases its error bar by 25\%. This lessens the tension with the  locally measured $H_0$  but it is insufficient to relieve it. However, an important feature of this model is to decouple the sound horizon drag scale at the recombination from the same scale in the local Universe. This  turns the BAO's metal fist delivering an inevitable knockout punch into a mere slap. We demonstrate this by considering a free dark energy equation of state. In the standard scenario, this model is very heavily degenerate using the CMB data alone, but the addition to BAO pins it back to the flat standard cosmology. In our case, by explicitly leaving the local baryon density as a free parameter subject to being drawn from the same distribution as those present on the surface of last scattering, the BAO does manage to tighten constraints, but the derived $H_0$ values easily accommodate the local measurements of the Hubble parameter. The price one has to pay, however, is deviation from $\Lambda$CDM as this model requires phantom dark energy to be consistent with both CMB and local measurements of $H_0$. You win some, you lose some. 

We stress that we use wCDM to illustrate that other approaches to solving the $H_0$ tension, that are often slaughtered by the BAO constraints, might get a second chance in this scenario. A careful assessment of these possibilities would require use of additional data (most notably supernovae Ia) and exceeds the scope of this paper.

\emph{Implications for the $S_8$ tension.} We have also found that $S_8$ naturally drops by around 1 sigma and its error bar increases by about 15\% (factor of 2 in the wCDM case), lessening the tension with the low redshift lensing measurements. This is different from, for example, early dark energy explanations of $H_0$ tension, which make this tension worse.  This is similar to the directions these parameters move when $A_L$ is varied, so it is possible that $S_8$ is in fact lower and that there exists an unknown systematic in the Planck data that smooths the peaks, as previously found in the Planck 2018 analysis.

\emph{Potential problems.} There are potential problems with this model. First, derivations of the $H_0$ parameter that rely on the matter-equality scale, or more specifically on the slope of the matter power spectrum \cite{2204.02984} are immune to variation in the sound horizon scale and also seem to prefer the low $H_0$. However, they hide other model assumptions that would need to be carefully evaluated (\cite{2208.12992}) Our model also might also have problems with the inverse distance ladder arguments that use Big Bang nucleosynthesis (BBN) determined $\omega_b$ \cite{1906.11628}, since these presumably rely on the local baryon density. We also note that the lensing reconstructions do not detect very large amounts of lensing power on the very large scales as one might naively expect, but these details depend on the details of the reconstruction algorithm and scales involved (our model predicts fluctuations on very large harmonic scales $\ell \lesssim 10$). 

\emph{Other predictions.} Our model makes certain other easily testable predictions. First, there should be an optimal way to detect the effect of the local power spectrum modulation from the four-point function of the CMB. Given the large scale cosmic variance, it is unlikely that this detection would be of a very high significance. However, variations in the sound horizon scale could be detected at lower redshift by considering variance in BAO parameters measured over patches of the sky at high redshifts or again a more optimally constructed four-point function. These tests could be performed by the proposed Stage 5 spectroscopic facility \cite{2209.03585}. Moreover, we expect that $B$-mode polarization would be generated in our model \cite{0907.3919}, albeit on very large scales. Detection of very large-scale B-mode polarization is therefore not a unique feature of the tensor fluctuations. However, B-mode polarization sourced by isocurvature would have particular correlations with temperature and $E$-mode polarization fluctuations that would allow us to disentangle it.

Another generic prediction of the baryon fraction modulation is that there will likely be a galaxy clustering bias associated with the local baryon density, i.e. $b_b = \partial \log \bar{n}_g / \partial \delta_{\obl}$, leading to large-scale modulation of galaxy clustering on the largest scales in the Universe. In fact, there are tentative detections of an excess of clustering of quasars and radio sources on the largest scales \cite{2009.14826,2206.05624} and associated theoretical developments \cite{2207.01569}. However, this is an extremely challenging measurement, especially relying on surveys that have not been explicitly designed with large-scale systematics control in mind. These should only be visible at $k \lesssim r_c^{-1} \sim  10^{-3}\,{\rm Mpc}^{-1}$, when relativistic effects starts to matter and where systematics render galaxy clustering measurements exceedingly difficult. Finally, this model also predicts large-scale correlations in the measurements of the baryon fraction by any astrophysical means (see, e.g., \cite{1208.5229,2013ApJ...778...14G,2023ApJ...944...50W}). Note that even if any one method is noisy and biased, owing to baryonic effects among others, there should be no large-scale correlations in any observable sensitive to a global baryon fraction in $\Lambda$CDM. Both of these approaches are difficult, since the signature will be present only on the largest scales accessible to cosmic surveys if $r_c$ is pushed to be large enough.

Finally, we would like to reiterate that the modeling in this paper was approximate. Correct modeling would involve exact evolution equations in the presence of very large superhorizon isocurvature fluctuations. These will predict the CMB power spectrum exactly and also predict the appropriate form of the CMB trispectrum sourced by them. 

\section*{Acknowledgments}

Authors thank Vivian Poulin for useful comments on the manuscript draft and thank Daniel Grin for useful discussions and pointing out previous work. We acknowledge discussions with Guillem Domenech and Eoin O Colgain.

AF acknowledges support from Brookhaven National Laboratory.  AS acknowledges hospitality of IFPU where this work has started. 


\section*{Appendix A: Probability for $\obl$}
\label{app:pdf}

Imagine drawing $N$ numbers $y_i$ from Gaussian distribution with a mean $m$ and variance $\sigma^2$. The probability for $m$ and $\sigma^2$ is given by the Bayesian theorem,
\begin{equation}
P(m, \sigma^2 | y_i) \propto \frac{1}{(2\pi \sigma^2)^{\sfrac{N}{2}}}\prod_{i=1}^{N} \exp\left( -\frac{(y_i-m)^2}{2 \sigma^2}\right).
\end{equation}
Defining the sample mean and variances as
\begin{eqnarray}
    m_s &=& \frac{1}{N} \sum_{i=1}^N y_i \\
    \sigma^2_s &=& = \frac{1}{N} \sum_{i=1}^N (y_i - m_s)^2 
\end{eqnarray}
we can rewrite
\begin{equation}
P(m, \sigma^2) = \frac{1}{(2\pi \sigma^2)^{\sfrac{N}{2}}} \exp \left(-\frac{N \left( \sigma_s^2-(m_s-m)^2\right)}{2 \sigma^2}\right).
\end{equation}
The probability of drawing the $(N+1)$-th number with a value $x$ is thus given by
\begin{equation}
     P(x) = \frac{1}{\sqrt{2\pi\sigma^2}} \iint \exp\left(\frac{-(x-m)^2}{2\sigma^2}\right) P(m, \sigma^2 | y_i) {\rm d}m\, {\rm d}\sigma^2
\end{equation}
Note that this probability corresponds to our subjective probability, given both the stochastic nature of the process as well as our imperfection in knowledge of true $\sigma^2$ and $m$ from a finite number of measurements $y_i$.  

Now we identify the $N$ numbers $y_i$ with the $N_p$ patches on the CMB sky that give some $\obc$ and $\sobc$ based on fitting the CMB peaks and their smoothness. Prior on our local baryon density $\obl$ is thus given by
\begin{widetext}
\begin{equation}
    P(\obl | \obc, \sobc) \propto \iint \frac{1}{(2\pi \sob^2)^{(N_p+1)/2}} \exp \left[-\frac{(\obl-\ob)^2 +N_p(\sobc^2+(\obc-\obl)^2) }{2\sob^2} \right] {\rm d}\ob\,{\rm d}\sob^2 \label{eq:oblprior}
\end{equation}
\end{widetext}
Integration over $\sigma_b$ can be done analytically, but integration of $\omega_b$ and determination of the normalization factor must be done analytically.

\section*{Appendix B: Triangle plots for non-minimal models.}
\label{app:triangles}
\begin{figure*}
    \centering
    \includegraphics[width=0.9\textwidth]{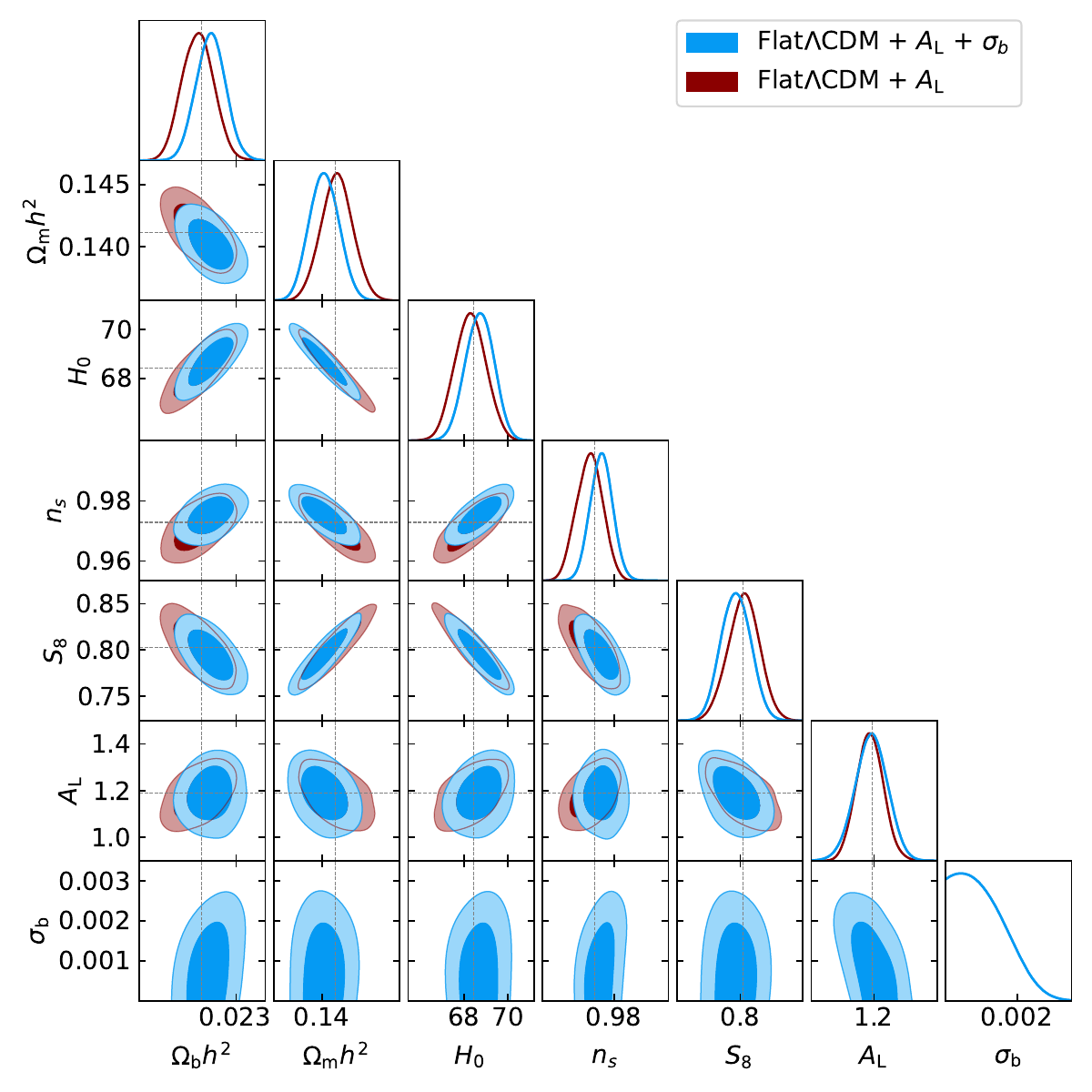}
    \caption{Corner plot for fits to Planck 2018 in the standard cosmology with $A_{\rm L}$ and one with large-scale baryon isocurvature fluctuations. Gray dashed lines are the best fit values from the standard case.}
    \label{fig:triangle_alens}
\end{figure*}

\begin{figure*}
    \centering
    \includegraphics[width=0.9\textwidth]{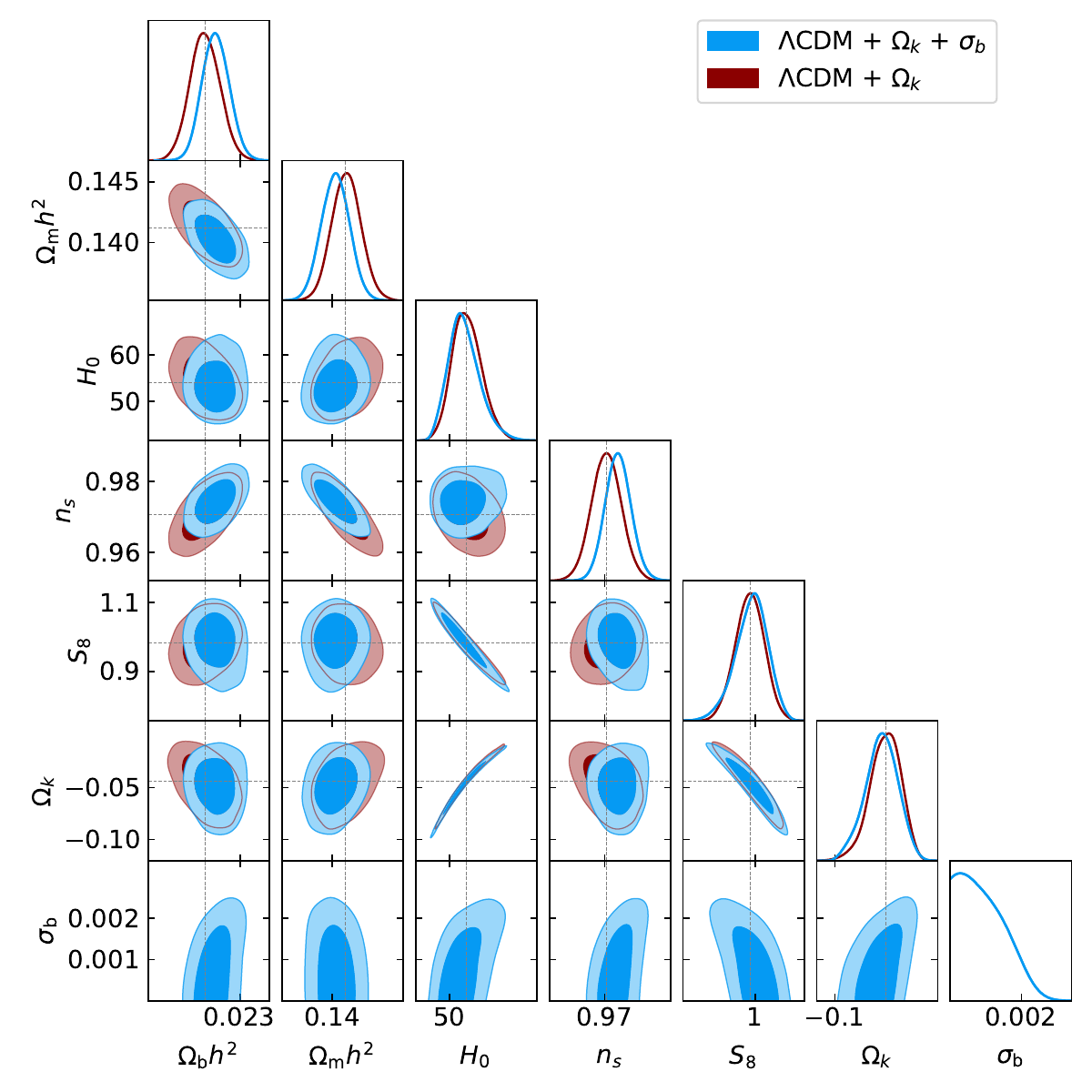}
    \caption{Corner plot for fits to Planck 2018 in the standard cosmology with $\Omega_k$ and one with large-scale baryon isocurvature fluctuations. Gray dashed lines are the best fit values from the standard case.}
    \label{fig:triangle_omegak}
\end{figure*}

\begin{figure*}
    \centering
    \includegraphics[width=0.9\textwidth]{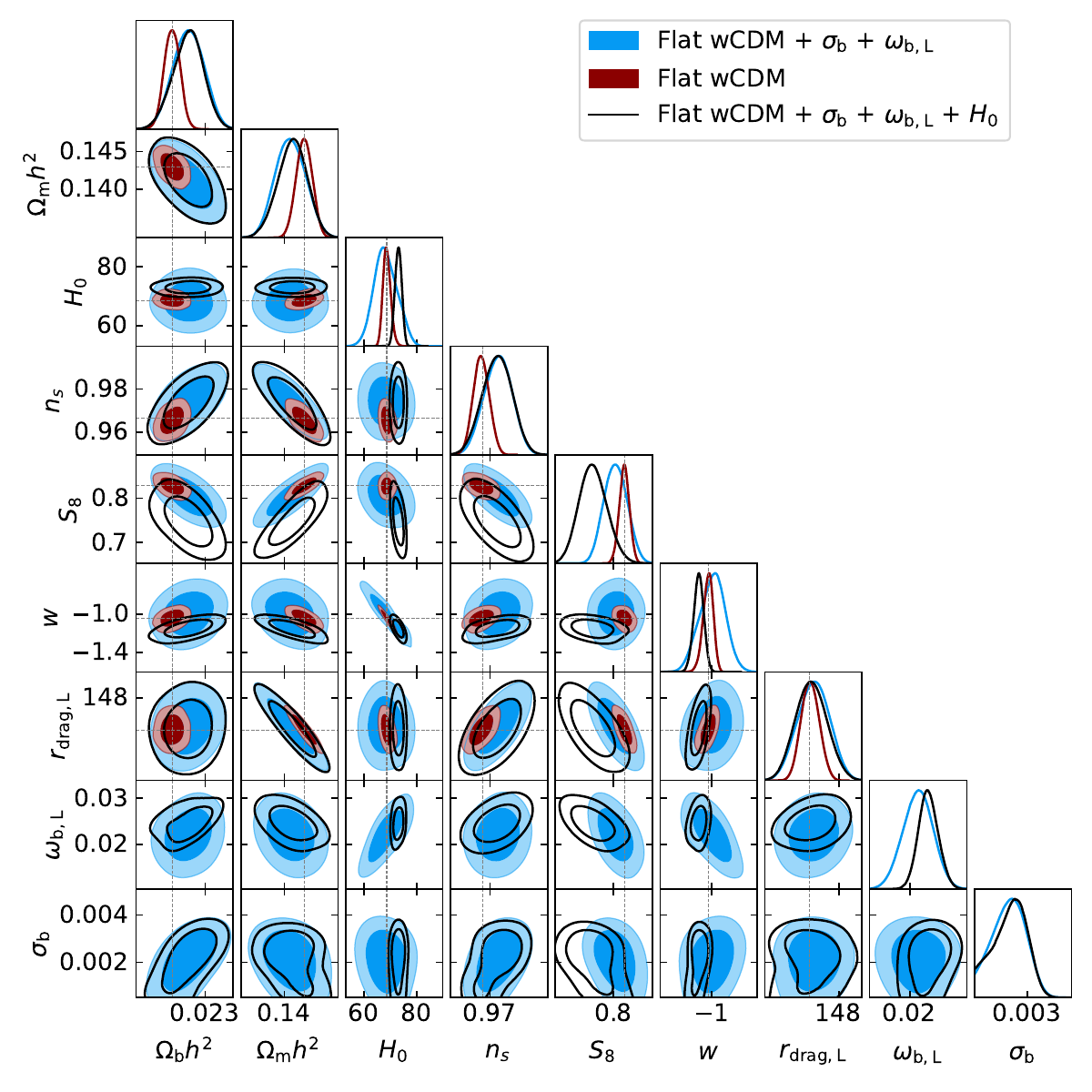}
    \caption{Corner plot for fits to Planck 2018 and BAO in the wCDM standard cosmology and one with large-scale baryon isocurvature fluctuations. Gray dashed lines are the best fit values from the standard case.}
    \label{fig:triangle_bao}
\end{figure*}

In this Appendix we show triangle plots for the $\Lambda$CDM model with addition of the lensing amplitude $A_L$ parameter (Fig. \ref{fig:triangle_alens}), the curvature parameter $\Omega_k$ (Fig. \ref{fig:triangle_omegak}), and the wCDM model where we have additionally used BAO data and importance sampled it with the distance ladder measurement (Fig. \ref{fig:triangle_bao}).

\bibliography{references.bib}

\begin{thebibliography}{37}%
\makeatletter
\providecommand \@ifxundefined [1]{%
 \@ifx{#1\undefined}
}%
\providecommand \@ifnum [1]{%
 \ifnum #1\expandafter \@firstoftwo
 \else \expandafter \@secondoftwo
 \fi
}%
\providecommand \@ifx [1]{%
 \ifx #1\expandafter \@firstoftwo
 \else \expandafter \@secondoftwo
 \fi
}%
\providecommand \natexlab [1]{#1}%
\providecommand \enquote  [1]{``#1''}%
\providecommand \bibnamefont  [1]{#1}%
\providecommand \bibfnamefont [1]{#1}%
\providecommand \citenamefont [1]{#1}%
\providecommand \href@noop [0]{\@secondoftwo}%
\providecommand \href [0]{\begingroup \@sanitize@url \@href}%
\providecommand \@href[1]{\@@startlink{#1}\@@href}%
\providecommand \@@href[1]{\endgroup#1\@@endlink}%
\providecommand \@sanitize@url [0]{\catcode `\\12\catcode `\$12\catcode
  `\&12\catcode `\#12\catcode `\^12\catcode `\_12\catcode `\%12\relax}%
\providecommand \@@startlink[1]{}%
\providecommand \@@endlink[0]{}%
\providecommand \url  [0]{\begingroup\@sanitize@url \@url }%
\providecommand \@url [1]{\endgroup\@href {#1}{\urlprefix }}%
\providecommand \urlprefix  [0]{URL }%
\providecommand \Eprint [0]{\href }%
\providecommand \doibase [0]{https://doi.org/}%
\providecommand \selectlanguage [0]{\@gobble}%
\providecommand \bibinfo  [0]{\@secondoftwo}%
\providecommand \bibfield  [0]{\@secondoftwo}%
\providecommand \translation [1]{[#1]}%
\providecommand \BibitemOpen [0]{}%
\providecommand \bibitemStop [0]{}%
\providecommand \bibitemNoStop [0]{.\EOS\space}%
\providecommand \EOS [0]{\spacefactor3000\relax}%
\providecommand \BibitemShut  [1]{\csname bibitem#1\endcsname}%
\let\auto@bib@innerbib\@empty
\bibitem [{\citenamefont {Riess}\ \emph {et~al.}(2022)\citenamefont {Riess}
  \emph {et~al.}}]{2112.04510}%
  \BibitemOpen
  \bibfield  {author} {\bibinfo {author} {\bibfnamefont {A.~G.}\ \bibnamefont
  {Riess}} \emph {et~al.},\ }\bibfield  {title} {\bibinfo {title} {{A
  Comprehensive Measurement of the Local Value of the Hubble Constant with 1 km
  s$^{-1}$ Mpc$^{-1}$ Uncertainty from the Hubble Space Telescope and the SH0ES
  Team}},\ }\href {https://doi.org/10.3847/2041-8213/ac5c5b} {\bibfield
  {journal} {\bibinfo  {journal} {Astrophys. J. Lett.}\ }\textbf {\bibinfo
  {volume} {934}},\ \bibinfo {pages} {L7} (\bibinfo {year} {2022})}\BibitemShut
  {NoStop}%
\bibitem [{\citenamefont {Aghanim}\ \emph {et~al.}(2020)\citenamefont {Aghanim}
  \emph {et~al.}}]{1807.06209}%
  \BibitemOpen
  \bibfield  {author} {\bibinfo {author} {\bibfnamefont {N.}~\bibnamefont
  {Aghanim}} \emph {et~al.} (\bibinfo {collaboration} {Planck Collaboration}),\
  }\bibfield  {title} {\bibinfo {title} {{Planck 2018 results. VI. Cosmological
  parameters}},\ }\href {https://doi.org/10.1051/0004-6361/201833910}
  {\bibfield  {journal} {\bibinfo  {journal} {Astron. Astrophys.}\ }\textbf
  {\bibinfo {volume} {641}},\ \bibinfo {pages} {A6} (\bibinfo {year} {2020})},\
  \bibinfo {note} {[Erratum: Astron.Astrophys. 652, C4 (2021)]}\BibitemShut
  {NoStop}%
\bibitem [{\citenamefont {Knox}\ and\ \citenamefont
  {Millea}(2020)}]{1908.03663}%
  \BibitemOpen
  \bibfield  {author} {\bibinfo {author} {\bibfnamefont {L.}~\bibnamefont
  {Knox}}\ and\ \bibinfo {author} {\bibfnamefont {M.}~\bibnamefont {Millea}},\
  }\bibfield  {title} {\bibinfo {title} {{Hubble constant
  hunter\textquoteright{}s guide}},\ }\href
  {https://doi.org/10.1103/PhysRevD.101.043533} {\bibfield  {journal} {\bibinfo
   {journal} {Phys. Rev. D}\ }\textbf {\bibinfo {volume} {101}},\ \bibinfo
  {pages} {043533} (\bibinfo {year} {2020})}\BibitemShut {NoStop}%
\bibitem [{\citenamefont {Sch\"oneberg}\ \emph {et~al.}(2022)\citenamefont
  {Sch\"oneberg}, \citenamefont {Franco~Abell\'an}, \citenamefont
  {P\'erez~S\'anchez}, \citenamefont {Witte}, \citenamefont {Poulin},\ and\
  \citenamefont {Lesgourgues}}]{2107.10291}%
  \BibitemOpen
  \bibfield  {author} {\bibinfo {author} {\bibfnamefont {N.}~\bibnamefont
  {Sch\"oneberg}}, \bibinfo {author} {\bibfnamefont {G.}~\bibnamefont
  {Franco~Abell\'an}}, \bibinfo {author} {\bibfnamefont {A.}~\bibnamefont
  {P\'erez~S\'anchez}}, \bibinfo {author} {\bibfnamefont {S.~J.}\ \bibnamefont
  {Witte}}, \bibinfo {author} {\bibfnamefont {V.}~\bibnamefont {Poulin}},\ and\
  \bibinfo {author} {\bibfnamefont {J.}~\bibnamefont {Lesgourgues}},\
  }\bibfield  {title} {\bibinfo {title} {{The H0 Olympics: A fair ranking of
  proposed models}},\ }\href {https://doi.org/10.1016/j.physrep.2022.07.001}
  {\bibfield  {journal} {\bibinfo  {journal} {Phys. Rept.}\ }\textbf {\bibinfo
  {volume} {984}},\ \bibinfo {pages} {1} (\bibinfo {year} {2022})}\BibitemShut
  {NoStop}%
\bibitem [{\citenamefont {Di~Valentino}\ \emph {et~al.}(2021)\citenamefont
  {Di~Valentino}, \citenamefont {Mena}, \citenamefont {Pan}, \citenamefont
  {Visinelli}, \citenamefont {Yang}, \citenamefont {Melchiorri}, \citenamefont
  {Mota}, \citenamefont {Riess},\ and\ \citenamefont {Silk}}]{2103.01183}%
  \BibitemOpen
  \bibfield  {author} {\bibinfo {author} {\bibfnamefont {E.}~\bibnamefont
  {Di~Valentino}}, \bibinfo {author} {\bibfnamefont {O.}~\bibnamefont {Mena}},
  \bibinfo {author} {\bibfnamefont {S.}~\bibnamefont {Pan}}, \bibinfo {author}
  {\bibfnamefont {L.}~\bibnamefont {Visinelli}}, \bibinfo {author}
  {\bibfnamefont {W.}~\bibnamefont {Yang}}, \bibinfo {author} {\bibfnamefont
  {A.}~\bibnamefont {Melchiorri}}, \bibinfo {author} {\bibfnamefont {D.~F.}\
  \bibnamefont {Mota}}, \bibinfo {author} {\bibfnamefont {A.~G.}\ \bibnamefont
  {Riess}},\ and\ \bibinfo {author} {\bibfnamefont {J.}~\bibnamefont {Silk}},\
  }\bibfield  {title} {\bibinfo {title} {{In the realm of the Hubble
  tension\textemdash{}a review of solutions}},\ }\href
  {https://doi.org/10.1088/1361-6382/ac086d} {\bibfield  {journal} {\bibinfo
  {journal} {Class. Quant. Grav.}\ }\textbf {\bibinfo {volume} {38}},\ \bibinfo
  {pages} {153001} (\bibinfo {year} {2021})}\BibitemShut {NoStop}%
\bibitem [{\citenamefont {Aubourg}\ \emph {et~al.}(2015)\citenamefont {Aubourg}
  \emph {et~al.}}]{1411.1074}%
  \BibitemOpen
  \bibfield  {author} {\bibinfo {author} {\bibfnamefont {E.}~\bibnamefont
  {Aubourg}} \emph {et~al.} (\bibinfo {collaboration} {BOSS Collaboration}),\
  }\bibfield  {title} {\bibinfo {title} {{Cosmological implications of baryon
  acoustic oscillation measurements}},\ }\href
  {https://doi.org/10.1103/PhysRevD.92.123516} {\bibfield  {journal} {\bibinfo
  {journal} {Phys. Rev. D}\ }\textbf {\bibinfo {volume} {92}},\ \bibinfo
  {pages} {123516} (\bibinfo {year} {2015})}\BibitemShut {NoStop}%
\bibitem [{\citenamefont {Cuesta}\ \emph {et~al.}(2015)\citenamefont {Cuesta},
  \citenamefont {Verde}, \citenamefont {Riess},\ and\ \citenamefont
  {Jimenez}}]{1411.1094}%
  \BibitemOpen
  \bibfield  {author} {\bibinfo {author} {\bibfnamefont {A.~J.}\ \bibnamefont
  {Cuesta}}, \bibinfo {author} {\bibfnamefont {L.}~\bibnamefont {Verde}},
  \bibinfo {author} {\bibfnamefont {A.}~\bibnamefont {Riess}},\ and\ \bibinfo
  {author} {\bibfnamefont {R.}~\bibnamefont {Jimenez}},\ }\bibfield  {title}
  {\bibinfo {title} {{Calibrating the cosmic distance scale ladder: the role of
  the sound horizon scale and the local expansion rate as distance anchors}},\
  }\href {https://doi.org/10.1093/mnras/stv261} {\bibfield  {journal} {\bibinfo
   {journal} {Mon. Not. Roy. Astron. Soc.}\ }\textbf {\bibinfo {volume}
  {448}},\ \bibinfo {pages} {3463} (\bibinfo {year} {2015})}\BibitemShut
  {NoStop}%
\bibitem [{\citenamefont {Holder}\ \emph {et~al.}(2010)\citenamefont {Holder},
  \citenamefont {Nollett},\ and\ \citenamefont {van Engelen}}]{0907.3919}%
  \BibitemOpen
  \bibfield  {author} {\bibinfo {author} {\bibfnamefont {G.~P.}\ \bibnamefont
  {Holder}}, \bibinfo {author} {\bibfnamefont {K.~M.}\ \bibnamefont
  {Nollett}},\ and\ \bibinfo {author} {\bibfnamefont {A.}~\bibnamefont {van
  Engelen}},\ }\bibfield  {title} {\bibinfo {title} {{On Possible Variation in
  the Cosmological Baryon Fraction}},\ }\href
  {https://doi.org/10.1088/0004-637X/716/2/907} {\bibfield  {journal} {\bibinfo
   {journal} {Astrophys. J.}\ }\textbf {\bibinfo {volume} {716}},\ \bibinfo
  {pages} {907} (\bibinfo {year} {2010})}\BibitemShut {NoStop}%
\bibitem [{\citenamefont {Grin}\ \emph {et~al.}(2014)\citenamefont {Grin},
  \citenamefont {Hanson}, \citenamefont {Holder}, \citenamefont {Dor\'e},\ and\
  \citenamefont {Kamionkowski}}]{1306.4319}%
  \BibitemOpen
  \bibfield  {author} {\bibinfo {author} {\bibfnamefont {D.}~\bibnamefont
  {Grin}}, \bibinfo {author} {\bibfnamefont {D.}~\bibnamefont {Hanson}},
  \bibinfo {author} {\bibfnamefont {G.~P.}\ \bibnamefont {Holder}}, \bibinfo
  {author} {\bibfnamefont {O.}~\bibnamefont {Dor\'e}},\ and\ \bibinfo {author}
  {\bibfnamefont {M.}~\bibnamefont {Kamionkowski}},\ }\bibfield  {title}
  {\bibinfo {title} {{Baryons do trace dark matter 380,000 years after the big
  bang: Search for compensated isocurvature perturbations with WMAP 9-year
  data}},\ }\href {https://doi.org/10.1103/PhysRevD.89.023006} {\bibfield
  {journal} {\bibinfo  {journal} {Phys. Rev. D}\ }\textbf {\bibinfo {volume}
  {89}},\ \bibinfo {pages} {023006} (\bibinfo {year} {2014})}\BibitemShut
  {NoStop}%
\bibitem [{\citenamefont {Smith}\ \emph {et~al.}(2017)\citenamefont {Smith},
  \citenamefont {Mu\~noz}, \citenamefont {Smith}, \citenamefont {Yee},\ and\
  \citenamefont {Grin}}]{1704.03461}%
  \BibitemOpen
  \bibfield  {author} {\bibinfo {author} {\bibfnamefont {T.~L.}\ \bibnamefont
  {Smith}}, \bibinfo {author} {\bibfnamefont {J.~B.}\ \bibnamefont {Mu\~noz}},
  \bibinfo {author} {\bibfnamefont {R.}~\bibnamefont {Smith}}, \bibinfo
  {author} {\bibfnamefont {K.}~\bibnamefont {Yee}},\ and\ \bibinfo {author}
  {\bibfnamefont {D.}~\bibnamefont {Grin}},\ }\bibfield  {title} {\bibinfo
  {title} {{Baryons still trace dark matter: probing CMB lensing maps for
  hidden isocurvature}},\ }\href {https://doi.org/10.1103/PhysRevD.96.083508}
  {\bibfield  {journal} {\bibinfo  {journal} {Phys. Rev. D}\ }\textbf {\bibinfo
  {volume} {96}},\ \bibinfo {pages} {083508} (\bibinfo {year}
  {2017})}\BibitemShut {NoStop}%
\bibitem [{\citenamefont {Ade}\ \emph {et~al.}(2016{\natexlab{a}})\citenamefont
  {Ade} \emph {et~al.}}]{1502.01589}%
  \BibitemOpen
  \bibfield  {author} {\bibinfo {author} {\bibfnamefont {P.~A.~R.}\
  \bibnamefont {Ade}} \emph {et~al.} (\bibinfo {collaboration} {Planck
  Collaboration}),\ }\bibfield  {title} {\bibinfo {title} {{Planck 2015
  results. XIII. Cosmological parameters}},\ }\href
  {https://doi.org/10.1051/0004-6361/201525830} {\bibfield  {journal} {\bibinfo
   {journal} {Astron. Astrophys.}\ }\textbf {\bibinfo {volume} {594}},\
  \bibinfo {pages} {A13} (\bibinfo {year} {2016}{\natexlab{a}})}\BibitemShut
  {NoStop}%
\bibitem [{\citenamefont {Di~Valentino}\ \emph {et~al.}(2019)\citenamefont
  {Di~Valentino}, \citenamefont {Melchiorri},\ and\ \citenamefont
  {Silk}}]{1911.02087}%
  \BibitemOpen
  \bibfield  {author} {\bibinfo {author} {\bibfnamefont {E.}~\bibnamefont
  {Di~Valentino}}, \bibinfo {author} {\bibfnamefont {A.}~\bibnamefont
  {Melchiorri}},\ and\ \bibinfo {author} {\bibfnamefont {J.}~\bibnamefont
  {Silk}},\ }\bibfield  {title} {\bibinfo {title} {{Planck evidence for a
  closed Universe and a possible crisis for cosmology}},\ }\href
  {https://doi.org/10.1038/s41550-019-0906-9} {\bibfield  {journal} {\bibinfo
  {journal} {Nature Astron.}\ }\textbf {\bibinfo {volume} {4}},\ \bibinfo
  {pages} {196} (\bibinfo {year} {2019})}\BibitemShut {NoStop}%
\bibitem [{\citenamefont {Handley}(2021)}]{1908.09139}%
  \BibitemOpen
  \bibfield  {author} {\bibinfo {author} {\bibfnamefont {W.}~\bibnamefont
  {Handley}},\ }\bibfield  {title} {\bibinfo {title} {{Curvature tension:
  evidence for a closed universe}},\ }\href
  {https://doi.org/10.1103/PhysRevD.103.L041301} {\bibfield  {journal}
  {\bibinfo  {journal} {Phys. Rev. D}\ }\textbf {\bibinfo {volume} {103}},\
  \bibinfo {pages} {L041301} (\bibinfo {year} {2021})}\BibitemShut {NoStop}%
\bibitem [{\citenamefont {Hikage}\ \emph {et~al.}(2019)\citenamefont {Hikage}
  \emph {et~al.}}]{1809.09148}%
  \BibitemOpen
  \bibfield  {author} {\bibinfo {author} {\bibfnamefont {C.}~\bibnamefont
  {Hikage}} \emph {et~al.} (\bibinfo {collaboration} {HSC Collaboration}),\
  }\bibfield  {title} {\bibinfo {title} {{Cosmology from cosmic shear power
  spectra with Subaru Hyper Suprime-Cam first-year data}},\ }\href
  {https://doi.org/10.1093/pasj/psz010} {\bibfield  {journal} {\bibinfo
  {journal} {Publ. Astron. Soc. Jap.}\ }\textbf {\bibinfo {volume} {71}},\
  \bibinfo {pages} {43} (\bibinfo {year} {2019})}\BibitemShut {NoStop}%
\bibitem [{\citenamefont {Abbott}\ \emph {et~al.}(2022)\citenamefont {Abbott}
  \emph {et~al.}}]{2105.13549}%
  \BibitemOpen
  \bibfield  {author} {\bibinfo {author} {\bibfnamefont {T.~M.~C.}\
  \bibnamefont {Abbott}} \emph {et~al.} (\bibinfo {collaboration} {DES
  Collaboration}),\ }\bibfield  {title} {\bibinfo {title} {{Dark Energy Survey
  Year 3 results: Cosmological constraints from galaxy clustering and weak
  lensing}},\ }\href {https://doi.org/10.1103/PhysRevD.105.023520} {\bibfield
  {journal} {\bibinfo  {journal} {Phys. Rev. D}\ }\textbf {\bibinfo {volume}
  {105}},\ \bibinfo {pages} {023520} (\bibinfo {year} {2022})}\BibitemShut
  {NoStop}%
\bibitem [{\citenamefont {Amon}\ \emph {et~al.}(2022)\citenamefont {Amon} \emph
  {et~al.}}]{2105.13543}%
  \BibitemOpen
  \bibfield  {author} {\bibinfo {author} {\bibfnamefont {A.}~\bibnamefont
  {Amon}} \emph {et~al.} (\bibinfo {collaboration} {DES Collaboration}),\
  }\bibfield  {title} {\bibinfo {title} {{Dark Energy Survey Year 3 results:
  Cosmology from cosmic shear and robustness to data calibration}},\ }\href
  {https://doi.org/10.1103/PhysRevD.105.023514} {\bibfield  {journal} {\bibinfo
   {journal} {Phys. Rev. D}\ }\textbf {\bibinfo {volume} {105}},\ \bibinfo
  {pages} {023514} (\bibinfo {year} {2022})}\BibitemShut {NoStop}%
\bibitem [{\citenamefont {Secco}\ \emph {et~al.}(2022)\citenamefont {Secco}
  \emph {et~al.}}]{2105.13544}%
  \BibitemOpen
  \bibfield  {author} {\bibinfo {author} {\bibfnamefont {L.~F.}\ \bibnamefont
  {Secco}} \emph {et~al.} (\bibinfo {collaboration} {DES Collaboration}),\
  }\bibfield  {title} {\bibinfo {title} {{Dark Energy Survey Year 3 results:
  Cosmology from cosmic shear and robustness to modeling uncertainty}},\ }\href
  {https://doi.org/10.1103/PhysRevD.105.023515} {\bibfield  {journal} {\bibinfo
   {journal} {Phys. Rev. D}\ }\textbf {\bibinfo {volume} {105}},\ \bibinfo
  {pages} {023515} (\bibinfo {year} {2022})}\BibitemShut {NoStop}%
\bibitem [{\citenamefont {Busch}\ \emph {et~al.}(2022)\citenamefont {Busch}
  \emph {et~al.}}]{2204.02396}%
  \BibitemOpen
  \bibfield  {author} {\bibinfo {author} {\bibfnamefont {J.~L. v.~d.}\
  \bibnamefont {Busch}} \emph {et~al.},\ }\bibfield  {title} {\bibinfo {title}
  {{KiDS-1000: Cosmic shear with enhanced redshift calibration}},\ }\href
  {https://doi.org/10.1051/0004-6361/202142083} {\bibfield  {journal} {\bibinfo
   {journal} {Astron. Astrophys.}\ }\textbf {\bibinfo {volume} {664}},\
  \bibinfo {pages} {A170} (\bibinfo {year} {2022})}\BibitemShut {NoStop}%
\bibitem [{\citenamefont {Ade}\ \emph {et~al.}(2016{\natexlab{b}})\citenamefont
  {Ade} \emph {et~al.}}]{1502.01597}%
  \BibitemOpen
  \bibfield  {author} {\bibinfo {author} {\bibfnamefont {P.~A.~R.}\
  \bibnamefont {Ade}} \emph {et~al.} (\bibinfo {collaboration} {Planck
  Collaboration}),\ }\bibfield  {title} {\bibinfo {title} {{Planck 2015
  results. XXIV. Cosmology from Sunyaev-Zeldovich cluster counts}},\ }\href
  {https://doi.org/10.1051/0004-6361/201525833} {\bibfield  {journal} {\bibinfo
   {journal} {Astron. Astrophys.}\ }\textbf {\bibinfo {volume} {594}},\
  \bibinfo {pages} {A24} (\bibinfo {year} {2016}{\natexlab{b}})}\BibitemShut
  {NoStop}%
\bibitem [{\citenamefont {Krolewski}\ \emph {et~al.}(2021)\citenamefont
  {Krolewski}, \citenamefont {Ferraro},\ and\ \citenamefont
  {White}}]{2105.03421}%
  \BibitemOpen
  \bibfield  {author} {\bibinfo {author} {\bibfnamefont {A.}~\bibnamefont
  {Krolewski}}, \bibinfo {author} {\bibfnamefont {S.}~\bibnamefont {Ferraro}},\
  and\ \bibinfo {author} {\bibfnamefont {M.}~\bibnamefont {White}},\ }\bibfield
   {title} {\bibinfo {title} {{Cosmological constraints from unWISE and Planck
  CMB lensing tomography}},\ }\href
  {https://doi.org/10.1088/1475-7516/2021/12/028} {\bibfield  {journal}
  {\bibinfo  {journal} {JCAP}\ }\textbf {\bibinfo {volume} {12}}\bibinfo
  {number} { (12)},\ \bibinfo {pages} {028}}\BibitemShut {NoStop}%
\bibitem [{\citenamefont {Abbott}\ \emph {et~al.}(2023)\citenamefont {Abbott}
  \emph {et~al.}}]{2206.10824}%
  \BibitemOpen
\bibfield  {number} {  }\bibfield  {author} {\bibinfo {author} {\bibfnamefont
  {T.~M.~C.}\ \bibnamefont {Abbott}} \emph {et~al.} (\bibinfo {collaboration}
  {DES Collaboration, SPT Collaboration}),\ }\bibfield  {title} {\bibinfo
  {title} {{Joint analysis of Dark Energy Survey Year 3 data and CMB lensing
  from SPT and Planck. III. Combined cosmological constraints}},\ }\href
  {https://doi.org/10.1103/PhysRevD.107.023531} {\bibfield  {journal} {\bibinfo
   {journal} {Phys. Rev. D}\ }\textbf {\bibinfo {volume} {107}},\ \bibinfo
  {pages} {023531} (\bibinfo {year} {2023})}\BibitemShut {NoStop}%
\bibitem [{\citenamefont {Heymans}\ \emph {et~al.}(2021)\citenamefont {Heymans}
  \emph {et~al.}}]{2007.15632}%
  \BibitemOpen
  \bibfield  {author} {\bibinfo {author} {\bibfnamefont {C.}~\bibnamefont
  {Heymans}} \emph {et~al.},\ }\bibfield  {title} {\bibinfo {title} {{KiDS-1000
  Cosmology: Multi-probe weak gravitational lensing and spectroscopic galaxy
  clustering constraints}},\ }\href
  {https://doi.org/10.1051/0004-6361/202039063} {\bibfield  {journal} {\bibinfo
   {journal} {Astron. Astrophys.}\ }\textbf {\bibinfo {volume} {646}},\
  \bibinfo {pages} {A140} (\bibinfo {year} {2021})}\BibitemShut {NoStop}%
\bibitem [{\citenamefont {Garc\'\i{}a-Garc\'\i{}a}\ \emph
  {et~al.}(2021)\citenamefont {Garc\'\i{}a-Garc\'\i{}a}, \citenamefont
  {Zapatero}, \citenamefont {Alonso}, \citenamefont {Bellini}, \citenamefont
  {Ferreira}, \citenamefont {Mueller}, \citenamefont {Nicola},\ and\
  \citenamefont {Ruiz-Lapuente}}]{2105.12108}%
  \BibitemOpen
  \bibfield  {author} {\bibinfo {author} {\bibfnamefont {C.}~\bibnamefont
  {Garc\'\i{}a-Garc\'\i{}a}}, \bibinfo {author} {\bibfnamefont {J.~R.}\
  \bibnamefont {Zapatero}}, \bibinfo {author} {\bibfnamefont {D.}~\bibnamefont
  {Alonso}}, \bibinfo {author} {\bibfnamefont {E.}~\bibnamefont {Bellini}},
  \bibinfo {author} {\bibfnamefont {P.~G.}\ \bibnamefont {Ferreira}}, \bibinfo
  {author} {\bibfnamefont {E.-M.}\ \bibnamefont {Mueller}}, \bibinfo {author}
  {\bibfnamefont {A.}~\bibnamefont {Nicola}},\ and\ \bibinfo {author}
  {\bibfnamefont {P.}~\bibnamefont {Ruiz-Lapuente}},\ }\bibfield  {title}
  {\bibinfo {title} {{The growth of density perturbations in the last
  \ensuremath{\sim}10 billion years from tomographic large-scale structure
  data}},\ }\href {https://doi.org/10.1088/1475-7516/2021/10/030} {\bibfield
  {journal} {\bibinfo  {journal} {JCAP}\ }\textbf {\bibinfo {volume} {10}},\
  \bibinfo {pages} {030}}\BibitemShut {NoStop}%
\bibitem [{\citenamefont {{Torrado}}\ and\ \citenamefont
  {{Lewis}}(2021)}]{Torrado:2020dgo}%
  \BibitemOpen
  \bibfield  {author} {\bibinfo {author} {\bibfnamefont {J.}~\bibnamefont
  {{Torrado}}}\ and\ \bibinfo {author} {\bibfnamefont {A.}~\bibnamefont
  {{Lewis}}},\ }\bibfield  {title} {\bibinfo {title} {{Cobaya: code for
  Bayesian analysis of hierarchical physical models}},\ }\href
  {https://doi.org/10.1088/1475-7516/2021/05/057} {\bibfield  {journal}
  {\bibinfo  {journal} {\jcap}\ }\textbf {\bibinfo {volume} {2021}},\ \bibinfo
  {eid} {057} (\bibinfo {year} {2021})},\ \Eprint
  {https://arxiv.org/abs/2005.05290} {arXiv:2005.05290 [astro-ph.IM]}
  \BibitemShut {NoStop}%
\bibitem [{\citenamefont {{Handley}}\ \emph {et~al.}(2015)\citenamefont
  {{Handley}}, \citenamefont {{Hobson}},\ and\ \citenamefont
  {{Lasenby}}}]{Handley:2015fda}%
  \BibitemOpen
  \bibfield  {author} {\bibinfo {author} {\bibfnamefont {W.~J.}\ \bibnamefont
  {{Handley}}}, \bibinfo {author} {\bibfnamefont {M.~P.}\ \bibnamefont
  {{Hobson}}},\ and\ \bibinfo {author} {\bibfnamefont {A.~N.}\ \bibnamefont
  {{Lasenby}}},\ }\bibfield  {title} {\bibinfo {title} {{polychord: nested
  sampling for cosmology.}},\ }\href {https://doi.org/10.1093/mnrasl/slv047}
  {\bibfield  {journal} {\bibinfo  {journal} {\mnras}\ }\textbf {\bibinfo
  {volume} {450}},\ \bibinfo {pages} {L61} (\bibinfo {year} {2015})},\ \Eprint
  {https://arxiv.org/abs/1502.01856} {arXiv:1502.01856 [astro-ph.CO]}
  \BibitemShut {NoStop}%
\bibitem [{\citenamefont {{Lewis}}\ \emph {et~al.}(2000)\citenamefont
  {{Lewis}}, \citenamefont {{Challinor}},\ and\ \citenamefont
  {{Lasenby}}}]{Lewis:1999bs}%
  \BibitemOpen
  \bibfield  {author} {\bibinfo {author} {\bibfnamefont {A.}~\bibnamefont
  {{Lewis}}}, \bibinfo {author} {\bibfnamefont {A.}~\bibnamefont
  {{Challinor}}},\ and\ \bibinfo {author} {\bibfnamefont {A.}~\bibnamefont
  {{Lasenby}}},\ }\bibfield  {title} {\bibinfo {title} {{Efficient Computation
  of Cosmic Microwave Background Anisotropies in Closed
  Friedmann-Robertson-Walker Models}},\ }\href {https://doi.org/10.1086/309179}
  {\bibfield  {journal} {\bibinfo  {journal} {\apj}\ }\textbf {\bibinfo
  {volume} {538}},\ \bibinfo {pages} {473} (\bibinfo {year} {2000})},\ \Eprint
  {https://arxiv.org/abs/astro-ph/9911177} {arXiv:astro-ph/9911177 [astro-ph]}
  \BibitemShut {NoStop}%
\bibitem [{\citenamefont {Alam}\ \emph {et~al.}(2017)\citenamefont {Alam} \emph
  {et~al.}}]{BOSS:2016wmc}%
  \BibitemOpen
  \bibfield  {author} {\bibinfo {author} {\bibfnamefont {S.}~\bibnamefont
  {Alam}} \emph {et~al.} (\bibinfo {collaboration} {BOSS Collaboration}),\
  }\bibfield  {title} {\bibinfo {title} {{The clustering of galaxies in the
  completed SDSS-III Baryon Oscillation Spectroscopic Survey: cosmological
  analysis of the DR12 galaxy sample}},\ }\href
  {https://doi.org/10.1093/mnras/stx721} {\bibfield  {journal} {\bibinfo
  {journal} {Mon. Not. Roy. Astron. Soc.}\ }\textbf {\bibinfo {volume} {470}},\
  \bibinfo {pages} {2617} (\bibinfo {year} {2017})}\BibitemShut {NoStop}%
\bibitem [{\citenamefont {Philcox}\ \emph {et~al.}(2022)\citenamefont
  {Philcox}, \citenamefont {Farren}, \citenamefont {Sherwin}, \citenamefont
  {Baxter},\ and\ \citenamefont {Brout}}]{2204.02984}%
  \BibitemOpen
  \bibfield  {author} {\bibinfo {author} {\bibfnamefont {O.~H.~E.}\
  \bibnamefont {Philcox}}, \bibinfo {author} {\bibfnamefont {G.~S.}\
  \bibnamefont {Farren}}, \bibinfo {author} {\bibfnamefont {B.~D.}\
  \bibnamefont {Sherwin}}, \bibinfo {author} {\bibfnamefont {E.~J.}\
  \bibnamefont {Baxter}},\ and\ \bibinfo {author} {\bibfnamefont {D.~J.}\
  \bibnamefont {Brout}},\ }\bibfield  {title} {\bibinfo {title} {{Determining
  the Hubble constant without the sound horizon: A 3.6\% constraint on H0 from
  galaxy surveys, CMB lensing, and supernovae}},\ }\href
  {https://doi.org/10.1103/PhysRevD.106.063530} {\bibfield  {journal} {\bibinfo
   {journal} {Phys. Rev. D}\ }\textbf {\bibinfo {volume} {106}},\ \bibinfo
  {pages} {063530} (\bibinfo {year} {2022})}\BibitemShut {NoStop}%
\bibitem [{\citenamefont {Smith}\ \emph {et~al.}(2023)\citenamefont {Smith},
  \citenamefont {Poulin},\ and\ \citenamefont {Simon}}]{2208.12992}%
  \BibitemOpen
  \bibfield  {author} {\bibinfo {author} {\bibfnamefont {T.~L.}\ \bibnamefont
  {Smith}}, \bibinfo {author} {\bibfnamefont {V.}~\bibnamefont {Poulin}},\ and\
  \bibinfo {author} {\bibfnamefont {T.}~\bibnamefont {Simon}},\ }\bibfield
  {title} {\bibinfo {title} {{Assessing the robustness of sound horizon-free
  determinations of the Hubble constant}},\ }\href
  {https://doi.org/10.1103/PhysRevD.108.103525} {\bibfield  {journal} {\bibinfo
   {journal} {Phys. Rev. D}\ }\textbf {\bibinfo {volume} {108}},\ \bibinfo
  {pages} {103525} (\bibinfo {year} {2023})}\BibitemShut {NoStop}%
\bibitem [{\citenamefont {Cuceu}\ \emph {et~al.}(2019)\citenamefont {Cuceu},
  \citenamefont {Farr}, \citenamefont {Lemos},\ and\ \citenamefont
  {Font-Ribera}}]{1906.11628}%
  \BibitemOpen
  \bibfield  {author} {\bibinfo {author} {\bibfnamefont {A.}~\bibnamefont
  {Cuceu}}, \bibinfo {author} {\bibfnamefont {J.}~\bibnamefont {Farr}},
  \bibinfo {author} {\bibfnamefont {P.}~\bibnamefont {Lemos}},\ and\ \bibinfo
  {author} {\bibfnamefont {A.}~\bibnamefont {Font-Ribera}},\ }\bibfield
  {title} {\bibinfo {title} {{Baryon Acoustic Oscillations and the Hubble
  Constant: Past, Present and Future}},\ }\href
  {https://doi.org/10.1088/1475-7516/2019/10/044} {\bibfield  {journal}
  {\bibinfo  {journal} {JCAP}\ }\textbf {\bibinfo {volume} {10}},\ \bibinfo
  {pages} {044}}\BibitemShut {NoStop}%
\bibitem [{\citenamefont {Schlegel}\ \emph {et~al.}(2022)\citenamefont
  {Schlegel} \emph {et~al.}}]{2209.03585}%
  \BibitemOpen
  \bibfield  {author} {\bibinfo {author} {\bibfnamefont {D.~J.}\ \bibnamefont
  {Schlegel}} \emph {et~al.} (\bibinfo {collaboration} {DESI Collaboration}),\
  }\bibfield  {title} {\bibinfo {title} {{A Spectroscopic Road Map for Cosmic
  Frontier: DESI, DESI-II, Stage-5}},\ }\href@noop {} {\  (\bibinfo {year}
  {2022})},\ \Eprint {https://arxiv.org/abs/2209.03585} {arXiv:2209.03585
  [astro-ph.CO]} \BibitemShut {NoStop}%
\bibitem [{\citenamefont {Secrest}\ \emph {et~al.}(2021)\citenamefont
  {Secrest}, \citenamefont {von Hausegger}, \citenamefont {Rameez},
  \citenamefont {Mohayaee}, \citenamefont {Sarkar},\ and\ \citenamefont
  {Colin}}]{2009.14826}%
  \BibitemOpen
  \bibfield  {author} {\bibinfo {author} {\bibfnamefont {N.~J.}\ \bibnamefont
  {Secrest}}, \bibinfo {author} {\bibfnamefont {S.}~\bibnamefont {von
  Hausegger}}, \bibinfo {author} {\bibfnamefont {M.}~\bibnamefont {Rameez}},
  \bibinfo {author} {\bibfnamefont {R.}~\bibnamefont {Mohayaee}}, \bibinfo
  {author} {\bibfnamefont {S.}~\bibnamefont {Sarkar}},\ and\ \bibinfo {author}
  {\bibfnamefont {J.}~\bibnamefont {Colin}},\ }\bibfield  {title} {\bibinfo
  {title} {{A Test of the Cosmological Principle with Quasars}},\ }\href
  {https://doi.org/10.3847/2041-8213/abdd40} {\bibfield  {journal} {\bibinfo
  {journal} {Astrophys. J. Lett.}\ }\textbf {\bibinfo {volume} {908}},\
  \bibinfo {pages} {L51} (\bibinfo {year} {2021})}\BibitemShut {NoStop}%
\bibitem [{\citenamefont {Secrest}\ \emph {et~al.}(2022)\citenamefont
  {Secrest}, \citenamefont {von Hausegger}, \citenamefont {Rameez},
  \citenamefont {Mohayaee},\ and\ \citenamefont {Sarkar}}]{2206.05624}%
  \BibitemOpen
  \bibfield  {author} {\bibinfo {author} {\bibfnamefont {N.~J.}\ \bibnamefont
  {Secrest}}, \bibinfo {author} {\bibfnamefont {S.}~\bibnamefont {von
  Hausegger}}, \bibinfo {author} {\bibfnamefont {M.}~\bibnamefont {Rameez}},
  \bibinfo {author} {\bibfnamefont {R.}~\bibnamefont {Mohayaee}},\ and\
  \bibinfo {author} {\bibfnamefont {S.}~\bibnamefont {Sarkar}},\ }\bibfield
  {title} {\bibinfo {title} {{A Challenge to the Standard Cosmological
  Model}},\ }\href {https://doi.org/10.3847/2041-8213/ac88c0} {\bibfield
  {journal} {\bibinfo  {journal} {Astrophys. J. Lett.}\ }\textbf {\bibinfo
  {volume} {937}},\ \bibinfo {pages} {L31} (\bibinfo {year}
  {2022})}\BibitemShut {NoStop}%
\bibitem [{\citenamefont {Dom\`enech}\ \emph {et~al.}(2022)\citenamefont
  {Dom\`enech}, \citenamefont {Mohayaee}, \citenamefont {Patil},\ and\
  \citenamefont {Sarkar}}]{2207.01569}%
  \BibitemOpen
  \bibfield  {author} {\bibinfo {author} {\bibfnamefont {G.}~\bibnamefont
  {Dom\`enech}}, \bibinfo {author} {\bibfnamefont {R.}~\bibnamefont
  {Mohayaee}}, \bibinfo {author} {\bibfnamefont {S.~P.}\ \bibnamefont
  {Patil}},\ and\ \bibinfo {author} {\bibfnamefont {S.}~\bibnamefont
  {Sarkar}},\ }\bibfield  {title} {\bibinfo {title} {{Galaxy number-count
  dipole and superhorizon fluctuations}},\ }\href
  {https://doi.org/10.1088/1475-7516/2022/10/019} {\bibfield  {journal}
  {\bibinfo  {journal} {JCAP}\ }\textbf {\bibinfo {volume} {10}},\ \bibinfo
  {pages} {019}}\BibitemShut {NoStop}%
\bibitem [{\citenamefont {Papastergis}\ \emph {et~al.}(2012)\citenamefont
  {Papastergis}, \citenamefont {Cattaneo}, \citenamefont {Huang}, \citenamefont
  {Giovanelli},\ and\ \citenamefont {Haynes}}]{1208.5229}%
  \BibitemOpen
  \bibfield  {author} {\bibinfo {author} {\bibfnamefont {E.}~\bibnamefont
  {Papastergis}}, \bibinfo {author} {\bibfnamefont {A.}~\bibnamefont
  {Cattaneo}}, \bibinfo {author} {\bibfnamefont {S.}~\bibnamefont {Huang}},
  \bibinfo {author} {\bibfnamefont {R.}~\bibnamefont {Giovanelli}},\ and\
  \bibinfo {author} {\bibfnamefont {M.~P.}\ \bibnamefont {Haynes}},\ }\bibfield
   {title} {\bibinfo {title} {{A direct measurement of the baryonic mass
  function of galaxies \& implications for the galactic baryon fraction}},\
  }\href {https://doi.org/10.1088/0004-637X/759/2/138} {\bibfield  {journal}
  {\bibinfo  {journal} {Astrophys. J.}\ }\textbf {\bibinfo {volume} {759}},\
  \bibinfo {pages} {138} (\bibinfo {year} {2012})}\BibitemShut {NoStop}%
\bibitem [{\citenamefont {Gonzalez}\ \emph {et~al.}(2013)\citenamefont
  {Gonzalez}, \citenamefont {Sivanandam}, \citenamefont {Zabludoff},\ and\
  \citenamefont {Zaritsky}}]{2013ApJ...778...14G}%
  \BibitemOpen
  \bibfield  {author} {\bibinfo {author} {\bibfnamefont {A.~H.}\ \bibnamefont
  {Gonzalez}}, \bibinfo {author} {\bibfnamefont {S.}~\bibnamefont
  {Sivanandam}}, \bibinfo {author} {\bibfnamefont {A.~I.}\ \bibnamefont
  {Zabludoff}},\ and\ \bibinfo {author} {\bibfnamefont {D.}~\bibnamefont
  {Zaritsky}},\ }\bibfield  {title} {\bibinfo {title} {{Galaxy cluster baryon
  fractions revisited}},\ }\href {https://doi.org/10.1088/0004-637X/778/1/14}
  {\bibfield  {journal} {\bibinfo  {journal} {Astrophys. J.}\ }\textbf
  {\bibinfo {volume} {778}},\ \bibinfo {pages} {14} (\bibinfo {year}
  {2013})}\BibitemShut {NoStop}%
\bibitem [{\citenamefont {Wang}\ and\ \citenamefont
  {Wei}(2023)}]{2023ApJ...944...50W}%
  \BibitemOpen
  \bibfield  {author} {\bibinfo {author} {\bibfnamefont {B.}~\bibnamefont
  {Wang}}\ and\ \bibinfo {author} {\bibfnamefont {J.-J.}\ \bibnamefont {Wei}},\
  }\bibfield  {title} {\bibinfo {title} {{An 8.0\% determination of the baryon
  fraction in the intergalactic medium from localized fast Radio Bursts}},\
  }\href {https://doi.org/10.3847/1538-4357/acb2c8} {\bibfield  {journal}
  {\bibinfo  {journal} {Astrophys. J.}\ }\textbf {\bibinfo {volume} {944}},\
  \bibinfo {pages} {50} (\bibinfo {year} {2023})}\BibitemShut {NoStop}%
\end{thebibliography}%

\end{document}